\documentclass[draft]{agujournal2019}

\usepackage{url}
\usepackage{lineno}
\usepackage[inline]{trackchanges}
\usepackage{soul}
\usepackage{amsmath,amssymb,array}
\usepackage{etoolbox}
\patchcmd{\correspondingauthor}{Corresponding author:}{Corresponding authors:}{}{}

\let\citep\shortcite

\draftfalse

\journalname{JGR: Solid Earth}

\begin{document}

\title{Efficient One-Step Surface-Wave Tomography through
Implicit Differentiation and Jacobian Factorization}

\authors{Yiran Jiang\affil{1}, Ping Tong\affil{2},
and Jianwei Ma\affil{3}}

\affiliation{1}{School of Mathematics, Harbin Institute of Technology,
China}
\affiliation{2}{Division of Mathematical Sciences, School of Physical and
Mathematical Sciences, Nanyang Technological University, Singapore}
\affiliation{3}{School of Earth and Space Sciences, Institute for Artificial
Intelligence, Peking University, China}

\correspondingauthor{Ping Tong and Jianwei Ma}{tongping@ntu.edu.sg; jwm@pku.edu.cn}

\begin{keypoints}
\item Implicit differentiation and reuse of the Eikonal dependency structure
efficiently construct individual traveltime sensitivity kernels
\item Jacobian factorization reduces storage and repeated Gauss--Newton
product costs by sharing model and dispersion mappings
\item Synthetic tests and 6.34 million U.S. traveltimes demonstrate
efficient one-step inversion of large surface-wave data sets
\end{keypoints}

\begin{abstract}
Surface-wave tomography provides important constraints on the crust and
upper mantle, but directly inverting millions of measurements for
three-dimensional structure remains computationally demanding. We develop
an efficient Gauss--Newton framework for one-step surface-wave tomography
that jointly inverts multiperiod interstation traveltimes for shear-wave
velocity, accounting for both lateral wave propagation and depth
sensitivity. Implicit differentiation of the converged discrete Eikonal
equations provides individual traveltime sensitivity kernels, which are
constructed efficiently by reusing the Eikonal dependency structure across
receivers. A factorized Jacobian retains these kernels separately from
shared model and dispersion mappings, reducing storage and repeated-product
costs within each linearized solve. Synthetic experiments
show that the framework recovers three-dimensional velocity
anomalies with fewer model updates and lower model errors than the tested
comparison workflows at similar data fits. Computational benchmarks
demonstrate substantial savings in operator storage and the cost of
linearized model updates compared with using explicitly assembled
Jacobians. We apply the framework to 6.34 million Rayleigh-wave phase
traveltimes across the contiguous United States. The recovered model
captures major crustal and uppermost mantle velocity variations, including
the broad contrast between the tectonically active western United States
and the continental interior, consistent with previous surface-wave
studies. These results demonstrate a practical Gauss--Newton approach to
imaging the crust and upper mantle using large surface-wave data sets.
\end{abstract}

\section*{Plain Language Summary}

Surface-wave observations provide a way to investigate the
three-dimensional structure of Earth's crust and uppermost mantle.
Measurements at different wave periods are sensitive to different depths,
while paths between recording stations sample lateral variations.
Combining these measurements directly allows their complementary
information to constrain a common three-dimensional model, rather than
first constructing separate maps at individual periods. However, doing so
with millions of observations requires substantial computing time and
memory. We develop an efficient method that calculates how individual
travel times respond to structural changes and reuses physical
relationships shared by many observations. This reduces repeated
calculations and storage as the model is updated. In synthetic tests, the
method recovers the known velocity structure with fewer updates and lower
model errors than the other tested inversion approaches at similar
agreement with the data. We apply the method to 6.34 million measurements
across the contiguous United States. The resulting model captures broad
crustal and uppermost mantle velocity variations, including slower wave
speeds beneath the tectonically active west and faster speeds beneath the
continental interior, consistent with previous studies. The approach makes
it more practical to translate large surface-wave data sets into
three-dimensional models of the crust and uppermost mantle.



\section{Introduction}
\label{sec:introduction}

Surface-wave dispersion provides important constraints on the shear-wave
velocity structure of the crust and uppermost mantle
\citep{RitzwollerLevshin1998,LinEtAl2009,ShenRitzwoller2016}. Measurements at
different periods are sensitive to different depth ranges, while the spatial
distribution of interstation paths provides lateral sampling. Dense regional
arrays and geographically extensive networks can both produce large
multiperiod surface-wave data sets. Combining their
complementary constraints in a three-dimensional model requires linking local
elastic structure to period-dependent surface-wave velocity and lateral
traveltime propagation. Efficiently incorporating these relationships into
the inversion is therefore important for translating large surface-wave data
sets into images of crustal and uppermost mantle structure.

Surface-wave observations are commonly interpreted through two sequential
inversions. Traveltimes are first inverted for two-dimensional velocity maps
at individual periods \citep{RitzwollerLevshin1998,BarminEtAl2001}. Local
dispersion curves extracted from these maps are then inverted for
one-dimensional shear-wave velocity profiles, which are assembled into a
three-dimensional model \citep{YaoEtAl2008}. Direct, or one-step,
formulations instead relate path-specific measurements to a common
three-dimensional model, retaining the coupling among periods, lateral
propagation, and depth sensitivity
\citep{BoschiEkstrom2002,FengAn2010,FangEtAl2015,ZhangEtAl2018MC,YangZhang2026}.
Local dispersion and traveltime propagation are updated together as the model
changes. However, constructing, storing, and repeatedly applying the
corresponding model-to-data Jacobian becomes costly as both the observation
count and the three-dimensional parameterization grow.

Adjoint-state methods provide an efficient way to use these sensitivities
without assembling the Jacobian when calculating a data-misfit gradient
\citep{Tarantola1984,Plessix2006}. Receiver residuals enter as adjoint
sources, allowing the contributions of individual observations to be combined
in an aggregate calculation. This approach has been developed for Eikonal
traveltime tomography \citep{LeungQian2006,TaillandierEtAl2009,Tong2021ATT}
and extended to surface-wave inversion
\citep{HaoEtAl2024,XuEtAl2025SurfATT,LeiEtAl2025}. The resulting gradient
incorporates the residual-weighted observation sensitivities and can be used
directly by gradient-based and quasi-Newton methods. These developments
provide the foundation for efficient sensitivity calculations in large
surface-wave inversions.

Gauss--Newton inversion uses the same first-order sensitivities to determine
a model update that jointly reduces the linearized traveltime residuals while
controlling the roughness and magnitude of the update. Its iterative
least-squares solution repeatedly applies the model-to-data derivative and
its transpose \citep{PaigeSaunders1982,FongSaunders2011}. This repeated use
creates an opportunity to reduce computation by constructing individual
traveltime sensitivity kernels once at the current model and retaining them
throughout the linearized solve. Once constructed, the kernels can
supply subsequent Jacobian products without the repeated tangent and adjoint
field calculations used in an on-demand implementation
\citep{MetivierEtAl2013}. Their construction cost can thus be distributed
over the many applications required for a model update. Realizing this
benefit for large surface-wave data sets requires both efficient construction
of individual traveltime sensitivity kernels and a compact representation of
their connection to three-dimensional structure.

Implicit differentiation of the converged discrete Eikonal equations provides
a route to traveltime sensitivity kernels. Previous studies have derived sensitivities directly from
discrete traveltime solvers \citep{LelievreEtAl2011}, including a factored
fast-marching formulation \citep{TreisterHaber2016} and a general discrete
Eikonal adjoint \citep{ZuninoEtAl2025}. These studies establish how the
numerical dependencies of the forward solution can be used in derivative
calculations. For the factored fast-sweeping solver considered here,
receivers associated with the same source and period share a converged
traveltime field and its linearized dependency structure. We exploit this
sharing by implicitly differentiating the converged equations and reusing
their dependency graph across receivers. The solution ordering is obtained
from the converged dependencies, including cycles, rather than assuming a
global acceptance order. Restricting each receiver-specific calculation to
the dependencies that influence that receiver then yields individual
traveltime sensitivity kernels without repeating a full-field adjoint calculation for
every observation.

The physical relationship between lateral propagation and local dispersion
provides a complementary way to retain these kernels economically. Each
traveltime kernel describes sensitivity to horizontal phase-slowness
variations at its period, whereas local dispersion kernels relate those
variations to the velocity profile beneath each horizontal node. The depth
mapping at a given node and period is shared by many observations. Explicitly
composing these relationships into three-dimensional sensitivity rows repeats
the same depth dependence for different paths, increasing storage and
repeated matrix-vector work. Retaining the observation-specific
phase-slowness kernels separately from the shared model and dispersion
mappings avoids this expansion. The resulting factorized Jacobian preserves
the complete model-to-data derivative while reducing the cost of storing and
repeatedly applying it. Thus, the factorization separates reusable
computational components without separating the inverse problem itself:
lateral propagation and local dispersion remain coupled, with all periods
continuing to constrain a common three-dimensional velocity model.

In this study, we combine implicit differentiation of the converged discrete
Eikonal equations with Jacobian factorization to develop an efficient
Gauss--Newton framework for one-step surface-wave tomography. The framework
extends SurfATT \citep{HaoEtAl2024,XuEtAl2025SurfATT}, reusing propagation
dependencies to construct individual traveltime sensitivity kernels
efficiently and retaining shared model and dispersion mappings to reduce
storage and repeated-product costs. We
verify the complete weighted derivative and its transpose, compare the
discrete kernels with continuous-adjoint calculations, and quantify the
computational benefits of kernel reuse and factorization. Synthetic
experiments evaluate model recovery and inversion cost relative to L-BFGS and
gradient-descent workflows. Finally, we apply the framework to 6.34 million
multiperiod Rayleigh-wave phase traveltimes across the contiguous United
States to demonstrate its practical use for imaging crustal and uppermost
mantle structure from a large observational data set.


\section{Methods}
\label{sec:methods}

\subsection{Forward Modeling and Inverse Formulation}
\label{sec:forward_problem}

\begin{figure}[t]
  \centering
  \includegraphics[width=\linewidth]{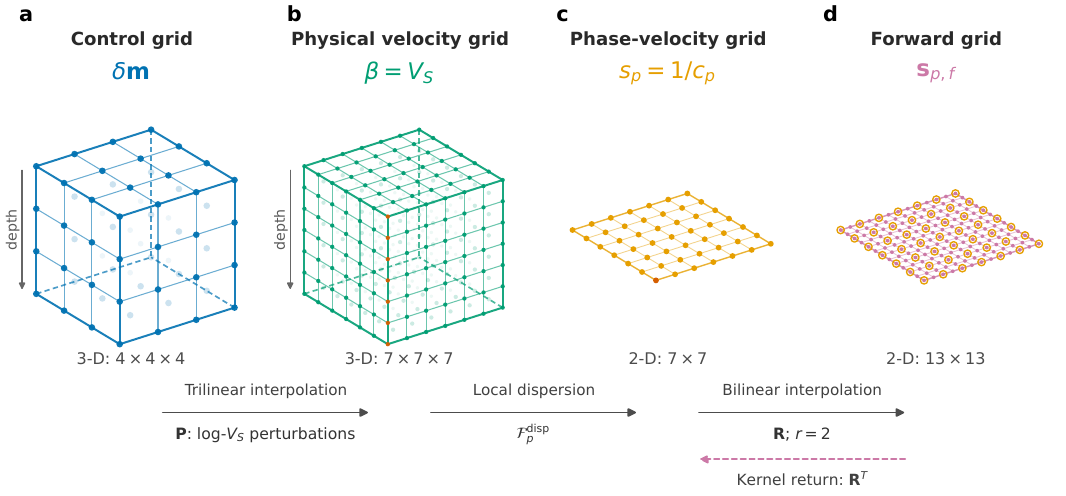}
  \caption{Grid hierarchy for one control grid and one period.
  (a,b) Trilinear interpolation $\mathbf P$ maps nodal log-velocity
  perturbations $\delta\mathbf m$ from the control grid to the
  three-dimensional physical grid carrying the current $V_S$ model.
  (c) Local dispersion followed by $s_p=1/c_p$ defines phase slowness at
  the physical grid's horizontal nodes.  The red edge-column nodes
  in (b) and node in (c) share the horizontal position $(x,y)=(0,0)$.
  (d) Bilinear interpolation $\mathbf R$ defines a finer two-dimensional
  forward grid; orange rings mark the original phase-velocity-grid nodes.
  The dashed arrow denotes sensitivity return by $\mathbf R^T$.
  Node counts are illustrative; $r=2$ is shown, whereas $r=1$ gives
  coincident phase-velocity and forward grids.}
  \label{fig:grid_hierarchy}
\end{figure}

Figure~\ref{fig:grid_hierarchy} summarizes the grid hierarchy used in the
forward and inverse problems.  Model changes on a sparse three-dimensional control grid
(Figure~\ref{fig:grid_hierarchy}a) are interpolated onto a denser physical
$V_S$ grid (Figure~\ref{fig:grid_hierarchy}b).  At each period, local
dispersion at its horizontal nodes gives phase velocity, whose reciprocal
defines phase slowness (Figure~\ref{fig:grid_hierarchy}c).
Bilinear interpolation gives the slowness on the finer forward grid used
for Eikonal propagation (Figure~\ref{fig:grid_hierarchy}d), and receiver
sampling gives the predicted interstation traveltimes.  A weighted
least-squares objective measures their differences from observations.  We
linearize these predictions with respect to control-grid changes and add
regularization to form the Gauss--Newton subproblem.

\subsubsection{Model parameterization and local dispersion}
\label{sec:parameterization_dispersion}

Surface waves are primarily sensitive to shear-wave speed
\citep{DahlenTromp1998}.  For the isotropic model considered here, we therefore
take $\beta=V_S$ as the independent variable and determine compressional-wave
speed $\alpha=V_P$ and density $\rho$ through prescribed relations
$\alpha(\beta,z)$ and $\rho(\beta,z)$, where $z$ is depth.
At a given iteration, let $\boldsymbol{\beta}$ denote the current
three-dimensional $V_S$ model on the physical grid
(Figure~\ref{fig:grid_hierarchy}b).  Each horizontal location on this grid
has a depth profile for the local dispersion calculation.  We parameterize
changes to $\boldsymbol{\beta}$ by nodal log-velocity perturbations
$\delta\mathbf{m}$ on the sparser control grid
(Figure~\ref{fig:grid_hierarchy}a).  Trilinear interpolation $\mathbf{P}$
maps these perturbations to the physical grid
(Figure~\ref{fig:grid_hierarchy}a,b):
\begin{equation}
  \delta\log\boldsymbol{\beta}=\mathbf{P}\,\delta\mathbf{m}.
  \label{eq:control_increment_mapping}
\end{equation}

The corresponding perturbation in shear-wave speed is
\begin{equation}
  \delta\boldsymbol{\beta}
  =\mathbf{D}\,\delta\log\boldsymbol{\beta}
  =\mathbf{D}\mathbf{P}\,\delta\mathbf{m},
  \qquad
  \mathbf{D}=\operatorname{diag}(\boldsymbol{\beta}),
  \label{eq:P_D}
\end{equation}

We assume that lateral variations are
smooth on the wavelength scale, allowing locally stratified dispersion
calculations and ray-theoretical propagation.  Let $\mathcal{P}$ be the set
of analyzed periods.  The phase-velocity grid consists of the horizontal
node locations of the physical grid (Figure~\ref{fig:grid_hierarchy}b,c).
At each node $\mathbf{x}$, layered-medium dispersion maps the local elastic
profiles to the phase velocity $c_p$ of the analyzed surface-wave mode at
period $p\in\mathcal{P}$
\citep{Thomson1950,Haskell1953,TakeuchiSaito1972}.  Denoting this nonlinear
operator by $\mathcal{F}^{\mathrm{disp}}_p$, we write
\begin{equation}
  c_p(\mathbf{x};\boldsymbol{\beta})
  =\mathcal{F}^{\mathrm{disp}}_p\!\left[
    \alpha\!\left(\beta(\mathbf{x},\cdot),\cdot\right),
    \beta(\mathbf{x},\cdot),
    \rho\!\left(\beta(\mathbf{x},\cdot),\cdot\right)
  \right],
  \label{eq:dispersion_forward}
\end{equation}
Here $\cdot$ denotes the depth coordinate along the full profile at fixed
$\mathbf{x}$; the prescribed relations are applied at each depth.
The phase slowness is
$s_p(\mathbf{x})=1/c_p(\mathbf{x})$, with nodal values $\mathbf{s}_p$
on the phase-velocity grid.

\subsubsection{Grid refinement and traveltime propagation}
\label{sec:propagation_refinement}

For each period, we interpolate phase slowness from the phase-velocity grid
to a finer forward grid used for traveltime propagation
(Figure~\ref{fig:grid_hierarchy}d).  A refinement factor $r$ subdivides each
phase-velocity-grid interval into $r$ forward-grid intervals.  Bilinear
interpolation gives
\begin{equation}
  \mathbf{s}_{p,f}=\mathbf{R}\mathbf{s}_p,
  \label{eq:slowness_grid_mapping}
\end{equation}
where $\mathbf{R}$ is the interpolation operator.  The grids coincide for
$r=1$, with $\mathbf{R}=\mathbf{I}$.

At each analyzed period, we treat a station as a virtual source and
calculate traveltimes to its receiver stations.  The observations for one
source and period form a \emph{gather} $g$, identified by source location
$\mathbf{x}_{s,g}$ and period $p_g\in\mathcal{P}$.  In the continuum, its
traveltime field $T_g$ satisfies the surface-wave Eikonal equation
\citep{Cerveny2001,LinEtAl2009}:
\begin{equation}
  \mathcal{E}_g(T_g,s_{p_g})
  = \frac{1}{2}\left[\nabla T_g\right]^{T}
    \mathbf{M}(\mathbf{x})\nabla T_g
    -\frac{1}{2}s_{p_g}^2(\mathbf{x})=0,
  \qquad
  T_g(\mathbf{x}_{s,g})=0,
  \label{eq:continuous_eikonal}
\end{equation}
Here $T_g(\mathbf{x})$ is the travel time from the source station to
$\mathbf{x}$, and $\mathbf{M}$ is the fixed inverse metric tensor of the
surface coordinates.  For a flat surface in Cartesian coordinates,
$\mathbf{M}$ reduces to the identity.  Spherical geometry enters through the
corresponding metric coefficients, and prescribed topography can be
incorporated in the same curvilinear representation \citep{HaoEtAl2024}.
We use $\mathbf{s}_{p_g,f}$ to solve the Eikonal equation on the forward
grid; the factored fast-sweeping formulation is described in
Section~\ref{sec:implicit_discrete_adjoint}.

\subsubsection{Traveltime data misfit}
\label{sec:receiver_misfit}

The predicted traveltime at a receiver is obtained by interpolation on the
forward grid.  Let $\mathcal{I}_g$ index the observations in gather $g$,
and let $\mathcal{R}_i$ denote sampling at receiver location
$\mathbf{x}_{r,i}$ for $i\in\mathcal{I}_g$.
Subtracting the observed traveltime $d_i^{\mathrm{obs}}$ from the prediction
$d_i$ gives the residual $e_i$:
\begin{equation}
  d_i(\boldsymbol{\beta})=\mathcal{R}_i T_g,
  \qquad
  e_i(\boldsymbol{\beta})=d_i(\boldsymbol{\beta})-d_i^{\mathrm{obs}}.
  \label{eq:data_residual}
\end{equation}
The receiver evaluation $\mathcal R_i$ is abstract here: the inversions use
the $T_0\tau$ implementation in equation~\eqref{eq:t0_tau_receiver}, which
interpolates the factored field $\tau$ and evaluates $T_0$ at the receiver.
Let $\mathbf{d}$ and $\mathbf{d}^{\mathrm{obs}}$ stack the predicted and
observed traveltimes in corresponding order.  For nonnegative observation
weights $w_i$, let $\mathbf W=\operatorname{diag}_i(w_i)$.  The data-misfit
objective is
\begin{equation}
  \Phi_d(\boldsymbol{\beta})
  = \frac{1}{2}\,\mathbf{e}^{T}\mathbf{W}\mathbf{e},
  \qquad
  \mathbf{e}=\mathbf{d}(\boldsymbol{\beta})-\mathbf{d}^{\mathrm{obs}}.
  \label{eq:data_objective}
\end{equation}

\subsubsection{Regularized Gauss--Newton inversion}
\label{sec:gn_inversion}

At nonlinear iteration $k$, let $\boldsymbol{\beta}_k$ be the current model,
$\mathbf e_k=\mathbf d(\boldsymbol{\beta}_k)-\mathbf d^{\mathrm{obs}}$
its traveltime residual, and $\mathbf W_k$ the data-weight matrix used in
that update.  The model-to-data Jacobian $\mathbf J_k$ maps a
log-velocity increment $\Delta\mathbf m$ on the control grid to the
first-order change in predicted traveltimes.  The corresponding linearized
residual is $\mathbf e_k+\mathbf J_k\Delta\mathbf m$.  Substituting it into
equation~\eqref{eq:data_objective} gives the data term
$\frac12\|\mathbf W_k^{1/2}(\mathbf e_k+\mathbf J_k\Delta\mathbf m)\|_2^2$.
We write $\mathbf J_{d,k}=\mathbf W_k^{1/2}\mathbf J_k$.
Constructing $\mathbf J_k$ for many observations is computationally costly;
Sections~\ref{sec:factorized_jacobian} and~\ref{sec:observation_jacobian}
describe its factorized representation and discrete construction.

We add roughness and damping penalties to this data term.  The operators
$\mathbf R_h$ and $\mathbf R_v$ penalize second-order horizontal and
vertical roughness of the physical $V_S$ increment
$\mathbf D_k\mathbf P\Delta\mathbf m$; $\mathbf R_0$ damps the physical-grid
log-$V_S$ increment $\mathbf P\Delta\mathbf m$.  Their scales are
$\gamma_{h,k}$, $\gamma_{v,k}$, and $\gamma_{0,k}$, respectively.
The resulting regularized Gauss--Newton subproblem, in which roughness and
damping act on the current update, is
\begin{equation}
  \begin{aligned}
  \Delta\mathbf{m}_k
  =\arg\min_{\Delta\mathbf{m}}
  &\ \frac{1}{2}\left\|
    \mathbf{J}_{d,k}\Delta\mathbf{m}
    +\mathbf{W}_k^{1/2}\mathbf{e}_k
  \right\|_2^2 \\
  &+\frac{1}{2}\left\|
    \gamma_{h,k}\mathbf{R}_h\mathbf{D}_k\mathbf{P}
    \Delta\mathbf{m}
  \right\|_2^2
  +\frac{1}{2}\left\|
    \gamma_{v,k}\mathbf{R}_v\mathbf{D}_k\mathbf{P}
    \Delta\mathbf{m}
  \right\|_2^2 \\
  &+\frac{1}{2}\left\|
    \gamma_{0,k}\mathbf{R}_0\mathbf{P}\Delta\mathbf{m}
  \right\|_2^2.
  \end{aligned}
  \label{eq:linearized_inverse}
\end{equation}
The regularization terms enter the augmented system as operator rows with
zero right-hand sides.  Their discrete rows, units, and boundary treatment
are specified in \ref{app:regularization_discretization}.

To specify the regularization levels, write
$\mathbf{L}_{h,k}=\mathbf{R}_h\mathbf{D}_k\mathbf{P}$,
$\mathbf{L}_{v,k}=\mathbf{R}_v\mathbf{D}_k\mathbf{P}$, and
$\mathbf{L}_{0,k}=\mathbf{R}_0\mathbf{P}$ for the operators with unit
regularization strength.  The dimensionless ratios $\theta_\ell$ set the
squared row weights:
\begin{equation}
  \gamma_{\ell,k}^{\,2}
  =\theta_\ell\,
    \max_{j\in\mathcal{S}_{\ell,k}}
    \frac{[\mathbf{J}_{d,k}^{T}\mathbf{J}_{d,k}]_{jj}}
         {[\mathbf{L}_{\ell,k}^{T}\mathbf{L}_{\ell,k}]_{jj}},
  \qquad \ell\in\{h,v,0\}.
  \label{eq:regularization_scaling}
\end{equation}
Here $j$ indexes control-grid parameters.  We include $j$ in
$\mathcal{S}_{\ell,k}$ when its diagonal value
$[\mathbf{L}_{\ell,k}^{T}\mathbf{L}_{\ell,k}]_{jj}$ exceeds $10^{-14}$
times the largest diagonal value of the same matrix.  This avoids a
near-zero denominator when setting $\gamma_{\ell,k}$; all control-grid
parameters remain in the inversion.  The largest ratio over this set
scales the regularization relative to the data term.  At each nonlinear
update, the diagonal of the weighted data normal matrix
$\mathbf J_{d,k}^T\mathbf J_{d,k}$ and the roughness diagonals are estimated
by probing, while the damping diagonal is computed directly
(\ref{app:regularization_discretization}).

We solve equation~\eqref{eq:linearized_inverse} with LSMR
\citep{FongSaunders2011}, applying the factorized Jacobian and
regularization operators through the augmented system.  The
individual traveltime sensitivity kernels are constructed once at
each nonlinear update and reused throughout the linear solve.

For one control grid, the physical-grid log-velocity direction is
$\Delta\mathbf u_k=\mathbf P\Delta\mathbf m_k$.
We cap the maximum absolute log-velocity increment at $b>0$ with
\[
  \kappa_k=\frac{b}{\max\{b,\|\Delta\mathbf u_k\|_\infty\}}.
\]

A single line search then tests the trial model
\begin{equation}
  \boldsymbol{\beta}^{\mathrm{trial}}
    =\boldsymbol{\beta}_k\odot
      \exp(\alpha\kappa_k\Delta\mathbf u_k),
  \label{eq:bounded_trial_update}
\end{equation}
where $\odot$ and the exponential act component-wise.  Backtracking starts
at $\alpha=1$; the accepted trial defines the next model.

For $n_{\mathrm{comp}}$ control grids, each solve uses the same residual
vector but its own interpolation $\mathbf P_c$ and regularization scales.
The coefficients are counted across grids, although the systems are
solved separately.  We average the resulting directions on the
physical grid \citep{TongEtAl2019MultipleGrid}:
\begin{equation}
  \Delta\mathbf u_k=\frac{1}{n_{\mathrm{comp}}}
    \sum_{c=1}^{n_{\mathrm{comp}}}\mathbf P_c\Delta\mathbf m_{c,k}.
  \label{eq:component_log_average}
\end{equation}
The same bound and a single line search are applied to this averaged
direction.

\subsection{Factorized Model-to-Data Jacobian}
\label{sec:factorized_jacobian}

Using the chain rule, we express the weighted model-to-data Jacobian as a
sequence of operators following the forward problem.  We then examine
how this Jacobian factorization can reduce storage and repeated Jacobian-product
work relative to explicit composition.

\subsubsection{Linearized operator chain}

For the model-to-data derivative, let
$\mathbf{c}=\operatorname{col}_{p\in\mathcal{P}}\mathbf{c}_p$ and
$\mathbf{s}=\operatorname{col}_{p\in\mathcal{P}}\mathbf{s}_p$ stack the
phase-velocity and phase-slowness fields over periods.  Likewise,
$\mathbf{s}_f=\operatorname{col}_{p\in\mathcal{P}}\mathbf{s}_{p,f}$ stacks
the forward-grid slownesses over periods, while
$\mathbf{T}=\operatorname{col}_g\mathbf{T}_g$ stacks the traveltime fields
over gathers.  The nonlinear model-to-data sequence is then
\begin{equation}
  \boldsymbol{\beta}
  \longmapsto \mathbf{c}
  \longmapsto \mathbf{s}
  \longmapsto \mathbf{s}_f
  \longmapsto \mathbf{T}
  \longmapsto \mathbf{d}.
  \label{eq:nonlinear_forward_chain}
\end{equation}
Source and receiver positions, period labels, the surface metric,
the prescribed elastic closure, and the data weights are held fixed
within each linearization.

Equation~\eqref{eq:P_D} first interpolates the control-grid log-velocity
perturbation $\delta\mathbf m$ with $\mathbf P$, then converts it to the
physical-grid shear-wave velocity perturbation $\delta\boldsymbol\beta$
with $\mathbf D=\operatorname{diag}(\boldsymbol\beta)$.
At each horizontal node, the phase-velocity
perturbation follows by differentiating local dispersion with respect
to the full-depth elastic profile.
Let $N_z^{\mathrm v}$ be the number of depth
nodes, and let $K_{p,k}^{\beta}$, $K_{p,k}^{\alpha}$, and
$K_{p,k}^{\rho}$ denote the partial derivatives of $c_p$ with respect to
$V_S$, $V_P$, and density at depth node $k$, respectively
\citep{TakeuchiSaito1972,DahlenTromp1998}.  Because $\alpha$ and $\rho$
follow the prescribed relations with $\beta$, the effective $V_S$
derivative is
\begin{equation}
  \delta c_p(\mathbf{x})
  =\sum_{k=1}^{N_z^{\mathrm v}}
  \left[
    K_{p,k}^{\beta}
    +K_{p,k}^{\alpha}
      \left.\frac{\partial\alpha}{\partial\beta}\right|_{z_k}
    +K_{p,k}^{\rho}
      \left.\frac{\partial\rho}{\partial\beta}\right|_{z_k}
  \right]_{\mathbf{x}}
  \delta\beta(\mathbf{x},z_k).
  \label{eq:dispersion_kernel}
\end{equation}
Stacking the effective kernels over horizontal position and period defines
$\mathbf{K}$, which is block diagonal with respect to horizontal position.
The coefficients include depth quadrature and the analytic derivatives of
the prescribed closure relations at fixed depth.
The conversion from phase velocity to phase slowness is diagonal,
\begin{equation}
  \begin{aligned}
    \delta s_p(\mathbf{x})
    &=-\frac{1}{c_p^2(\mathbf{x})}\,\delta c_p(\mathbf{x}),\\
    \delta\mathbf{s}
    &=\mathbf{Q}\,\delta\mathbf{c},
    \qquad
    \mathbf{Q}
    =\operatorname{diag}_{p,\mathbf{x}}
     \!\left[-c_p^{-2}(\mathbf{x})\right].
  \end{aligned}
  \label{eq:Q}
\end{equation}

The period-stacked slowness vector $\mathbf{s}$ is passed to the gathers
through $\mathbf{C}$, which selects the period field used by each gather and
places it in the gather layout $\mathbf{s}^{\mathrm{gath}}$.  For gather $g$,
define $\mathbf{A}_g$ as the derivative of its calculated traveltime vector
$\mathbf{d}_g$ with respect to the phase-slowness field $\mathbf{s}_{p_g}$:
\begin{equation}
  \delta\mathbf{d}_g=\mathbf{A}_g\delta\mathbf{s}_{p_g}.
  \label{eq:gather_A}
\end{equation}
This derivative includes interpolation to the forward grid, Eikonal
propagation, and receiver sampling.  In particular, its columns remain on
the phase-velocity grid when the forward grid is refined.  Assembling the
gather derivatives into $\mathbf{A}$ gives
\begin{equation}
  \mathbf{s}^{\mathrm{gath}}=\mathbf{C}\mathbf{s},
  \qquad
  \delta\mathbf{d}=\mathbf{A}\delta\mathbf{s}^{\mathrm{gath}}
  =\mathbf{A}\mathbf{C}\delta\mathbf{s}.
  \label{eq:gather_layout}
\end{equation}
Each row of $\mathbf{A}$ acts on its gather's phase-slowness copy.
Here $N_d$ is the number of observations, $N_p=|\mathcal{P}|$ is the number
of periods, $N_h^{\phi}$ is the number of phase-velocity nodes per period,
and $N_\ell$ counts phase-slowness values in the gather layout.  Accordingly,
$\mathbf{A}\in\mathbb{R}^{N_d\times N_\ell}$ and
$\mathbf{C}\in\mathbb{R}^{N_\ell\times N_pN_h^{\phi}}$.
The transpose $\mathbf{A}^T$ sums contributions within each gather, and
$\mathbf{C}^T$ accumulates them by period and node.

Combining these derivatives by the chain rule gives the weighted
Jacobian with respect to control-grid log-velocity perturbations:
\begin{equation}
  \boxed{
  \mathbf{J}_d
  =\mathbf{W}^{1/2}\mathbf{A}\mathbf{C}
   \mathbf{Q}\mathbf{K}\mathbf{D}\mathbf{P}.}
  \label{eq:factorized_J}
\end{equation}

In this factorization, $\mathbf A$ contains the observation-specific
lateral-propagation sensitivity kernels, whereas
$\mathbf C\mathbf Q\mathbf K\mathbf D\mathbf P$ contains the shared
mapping from control-grid log-$V_S$ perturbations to the gather-layout,
period-dependent phase slowness.  These factors are retained separately
for computation but remain coupled in every Jacobian product and
Gauss--Newton update.

Explicit composition expands $\mathbf A$ over depth, increasing storage
and matrix-vector work.  We retain its factors, applying them right to left
and their transposes in reverse:
\begin{equation}
  \mathbf{J}_d^T\mathbf{y}
  =\mathbf{P}^T\mathbf{D}^T\mathbf{K}^T\mathbf{Q}^T
   \mathbf{C}^T\mathbf{A}^T
   \left(\mathbf{W}^{1/2}\right)^T\mathbf{y}.
  \label{eq:factorized_transpose}
\end{equation}
The remaining observation-dependent factor is $\mathbf A$.
Section~\ref{sec:observation_jacobian} derives its individual rows from the
converged discrete Eikonal system.

\subsubsection{Storage and Matrix-Vector-Product Scaling}
\label{sec:storage_scaling}

We compare storage and matrix-vector work for the factorized and
explicitly composed Jacobian representations.  Let
$N_h^{\phi}=N_x^{\phi}N_y^{\phi}$ denote the number of horizontal
phase-velocity nodes for one period.  For one control grid, let
$N_h^{\mathrm m}$ and $N_z^{\mathrm m}$ denote the horizontal
and vertical numbers of seismic velocity controls, respectively, with
$N_c=N_h^{\mathrm m}N_z^{\mathrm m}$ parameters.  With $N_p=|\mathcal P|$ periods
and $N_d$ observations, the physical model and period-stacked slowness
field contain $N_v=N_h^{\phi}N_z^{\mathrm v}$ and
$N_s=N_h^{\phi}N_p$ values, respectively.

Let $\eta_h$ be the mean fraction of phase-velocity nodes represented in
an observation row.  Its mean coefficient count is
\begin{equation}
  \overline n_{\phi}
  =\frac{\operatorname{nnz}\mathbf{A}}{N_d}
  =\eta_hN_h^{\phi},
  \qquad
  \operatorname{nnz}\mathbf{A}=N_d\overline n_{\phi}.
  \label{eq:horizontal_sparsity}
\end{equation}
The layout map $\mathbf C$ uses $O(N_\ell)$ field workspace without a
stored sparse matrix.  With sparse $\mathbf P$, diagonal $\mathbf D$ and
$\mathbf Q$, and block-local $\mathbf K$, the factorized storage scales as
\begin{equation}
  O\!\left(
    \eta_hN_dN_h^{\phi}
    +\operatorname{nnz}\mathbf{P}+N_v
    +N_h^{\phi}N_pN_z^{\mathrm v}+N_s+N_\ell+N_d
  \right).
  \label{eq:factorized_storage}
\end{equation}
For fixed grids, periods, and layout, only the $\mathbf A$ rows and data
weights grow with the observation count.

After explicit composition, let $\eta_h^{\mathrm m}$ denote the mean
horizontal support fraction on the control grid.  If the dispersion
blocks have structural support over all $N_z^{\mathrm m}$ depth controls,
then
\begin{equation}
  \overline n_m
  =\frac{\operatorname{nnz}\mathbf J_d}{N_d}
  \simeq \eta_h^{\mathrm m}N_h^{\mathrm m}N_z^{\mathrm m},
  \qquad
  \operatorname{nnz}\mathbf J_d
  \simeq \eta_h^{\mathrm m}N_dN_h^{\mathrm m}N_z^{\mathrm m}.
  \label{eq:composite_cost}
\end{equation}
The ratio of observation-dependent coefficient counts is
\begin{equation}
  \frac{\operatorname{nnz}\mathbf J_d}
       {\operatorname{nnz}\mathbf A}
  \simeq
  \frac{\eta_h^{\mathrm m}}{\eta_h}
  \frac{N_c}{N_h^{\phi}},
  \qquad
  \frac{N_c}{N_h^{\phi}}=
  \frac{N_h^{\mathrm m}N_z^{\mathrm m}}{N_h^{\phi}}.
  \label{eq:factorization_reduction}
\end{equation}
Equation~\eqref{eq:factorization_reduction} compares the
observation-dependent coefficients stored and multiplied by the two
representations; a factorized product also applies the shared operators.
The grid-size factor
$N_c/N_h^{\phi}=(N_h^{\mathrm m}/N_h^{\phi})N_z^{\mathrm m}$
combines the number of depth controls with the relative horizontal grid
sizes.  If the horizontal node counts and support fractions are similar,
the composite rows have about $N_z^{\mathrm m}$ times as many
coefficients as the rows of $\mathbf A$.  At fixed control-grid size,
increasing the number of phase-velocity nodes lowers this factor, while
interpolation can enlarge the fraction of horizontal control nodes
touched by a row and raise $\eta_h^{\mathrm m}/\eta_h$.

For multiple control grids, cumulative explicit storage sums their
composed matrices, whereas
factorization shares $\mathbf A$, $\mathbf C$, $\mathbf Q$, $\mathbf K$,
$\mathbf D$, and $\mathbf W^{1/2}$ across grids and retains a separate
$\mathbf P$ for each grid.  Row-wise thresholding can further reduce
$\mathbf A$ when storage is limiting; its effect is assessed in
Section~\ref{sec:factorized_scaling_results}.

\subsection{Construction of Individual Traveltime Sensitivity Kernels}
\label{sec:observation_jacobian}

The factor $\mathbf{A}$ contains one phase-slowness sensitivity row per
observed traveltime.  This section describes how these rows are constructed
from the converged discrete Eikonal solution.  Within each gather, the
receivers share the same Eikonal field and linearized dependency structure,
allowing the gather-level quantities to be reused while the individual
traveltime sensitivity kernels are evaluated separately.

\subsubsection{Implicit differentiation and the discrete adjoint}
\label{sec:implicit_discrete_adjoint}

We first derive this shared linearized system and then specialize it to an
individual receiver.

We follow the factored fast-sweeping formulation used in SurfATT to
account for the point-source singularity
\citep{FomelLuoZhao2009,HaoEtAl2024,XuEtAl2025SurfATT}.
At forward-grid node $j$, the traveltime is written as
\begin{equation}
  T_{g,j}=T_{0,g,j}\tau_{g,j},
  \qquad
  T_{0,g,j}=d_{\mathcal{S}}(\mathbf{x}_j,\mathbf{x}_{s,g})
  s_{p_g,f}(\mathbf{x}_{s,g}),
  \label{eq:factored_field}
\end{equation}
where $d_{\mathcal{S}}$ is surface distance.  The source slowness is
$s_{p_g,f}(\mathbf{x}_{s,g})=\boldsymbol{\chi}_{s,g}^T\mathbf{s}_{p_g,f}$,
where $\boldsymbol{\chi}_{s,g}$ contains the bilinear interpolation weights
at the source.  At free nodes, $\tau_{g,j}$ is computed by alternating
Gauss--Seidel sweeps \citep{Zhao2005}.

Let $\boldsymbol{\tau}_g$ collect these free-node values $\tau_{g,j}$, and let
$\boldsymbol{\phi}_g(\boldsymbol{\tau}_g,\mathbf{s}_{p_g,f})$ denote the
collection of selected local nodal updates reconstructed from the converged
field, with their active branches held fixed during linearization
(\ref{app:stencil}).
At convergence, the field satisfies the fixed-point system
\begin{equation}
  \boldsymbol{\tau}_g
  =\boldsymbol{\phi}_g(\boldsymbol{\tau}_g,\mathbf{s}_{p_g,f}),
  \qquad
  \mathbf{F}_g
  =\boldsymbol{\tau}_g-
   \boldsymbol{\phi}_g(\boldsymbol{\tau}_g,\mathbf{s}_{p_g,f})=\mathbf{0}.
  \label{eq:discrete_fixed_point}
\end{equation}
Here $\mathbf{F}_g$ is the discrete Eikonal residual.  Differentiating the
update operator gives
\begin{equation}
  \mathbf{U}_{\tau,g}
  =\frac{\partial\boldsymbol{\phi}_g}
         {\partial\boldsymbol{\tau}_g},
  \qquad
  \mathbf{U}_{s,g}
  =\frac{\partial\boldsymbol{\phi}_g}{\partial\mathbf{s}_{p_g,f}},
  \qquad
  \mathbf{B}_g=\mathbf{I}-\mathbf{U}_{\tau,g}.
  \label{eq:active_matrices}
\end{equation}
Assuming $\mathbf{B}_g$ is nonsingular, implicit differentiation
\citep{GilesPierce2000} of
equation~\eqref{eq:discrete_fixed_point} gives the tangent equation
\begin{equation}
  \mathbf{B}_g\,\delta\boldsymbol{\tau}_g
  =\mathbf{U}_{s,g}\,\delta\mathbf{s}_{p_g,f}.
  \label{eq:discrete_tangent}
\end{equation}
A calculated traveltime depends on phase slowness indirectly through the
converged factored field $\boldsymbol\tau_g$ and explicitly through the
source factor $T_0$.  To keep these contributions separate, consider a
general differentiable scalar response
$\Psi_g(\boldsymbol\tau_g,\mathbf s_{p_g,f})$.  Treating its two arguments
as independent before enforcing the Eikonal constraint gives
\begin{equation}
  \delta\Psi_g
  =\left(\nabla_{\boldsymbol{\tau}_g}\Psi_g\right)^T
    \delta\boldsymbol{\tau}_g
   +\left(
      \left.\nabla_{\mathbf{s}_{p_g,f}}\Psi_g
      \right|_{\boldsymbol{\tau}_g}
    \right)^T\delta\mathbf{s}_{p_g,f}.
  \label{eq:general_scalar_variation}
\end{equation}
The first term contains the field-mediated dependence through
$\boldsymbol\tau_g$, whereas the second contains explicit slowness
dependence at fixed $\boldsymbol\tau_g$.  Defining the discrete adjoint by
\begin{equation}
  \mathbf{B}_g^T\boldsymbol{\lambda}_{\Psi,g}
  =\nabla_{\boldsymbol{\tau}_g}\Psi_g
  \label{eq:general_discrete_adjoint}
\end{equation}
and using equation~\eqref{eq:discrete_tangent} gives
\begin{equation}
  \nabla_{\mathbf{s}_{p_g,f}}\Psi_g
  =\left.\nabla_{\mathbf{s}_{p_g,f}}\Psi_g
    \right|_{\boldsymbol{\tau}_g}
   +\mathbf{U}_{s,g}^T\boldsymbol{\lambda}_{\Psi,g},
  \label{eq:general_discrete_gradient}
\end{equation}
Equation~\eqref{eq:general_discrete_gradient} is the common adjoint identity
used below for both an individual traveltime and the aggregate data-misfit
gradient.

\begin{figure}[t]
  \centering
  \includegraphics[width=\linewidth]{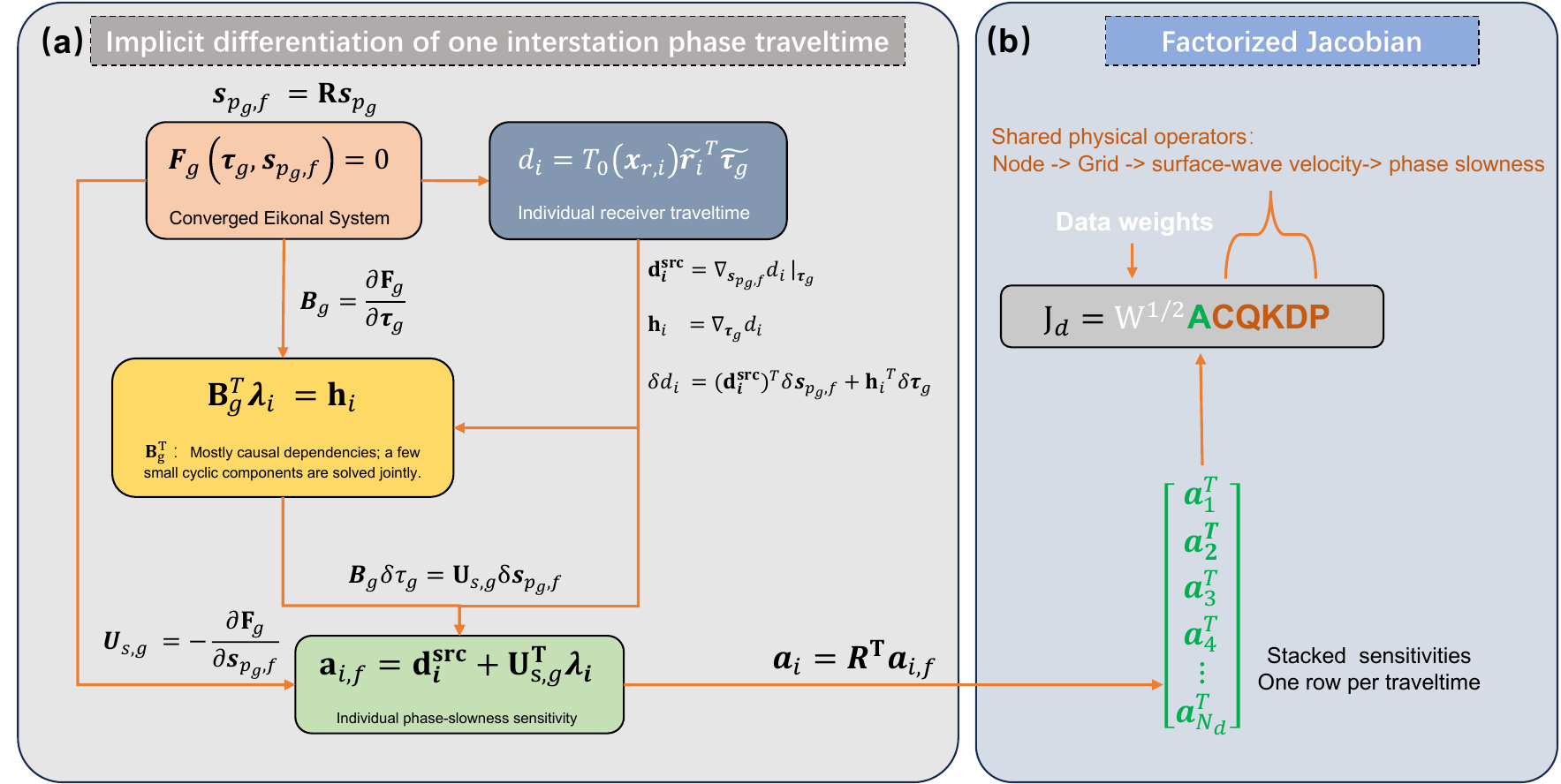}
  \caption{Summary of the construction of an individual traveltime sensitivity
  and the factorized model-to-data Jacobian.  (a) Period-dependent phase slowness is interpolated from the
  phase-velocity grid to the forward grid by $\mathbf R$.
  Gather $g$ uses the field $\mathbf s_{p_g,f}$.  For observation $i$ in this gather,
  the converged discrete
  Eikonal residual and receiver sampling define the derivatives
  $\mathbf B_g$, $\mathbf U_{s,g}$, $\mathbf h_i$, and
  $\mathbf d_i^{\mathrm{src}}$, with the active update branches held fixed.
  The tangent equation and the transpose solve
  $\mathbf B_g^T\boldsymbol\lambda_i=\mathbf h_i$ give the individual
  forward-grid sensitivity $\mathbf a_{i,f}$.  Acyclic dependencies are
  accumulated in reverse causal order, and cyclic components are solved
  jointly.  Transpose interpolation returns
  $\mathbf a_i=\mathbf R^T\mathbf a_{i,f}$ to the phase-velocity grid.
  (b) Placing each returned row $\mathbf a_i^T$ in its period-field columns
  in the gather layout forms $\mathbf A$, with one sparse row per traveltime.
  Combining $\mathbf A$ with
  the data weights $\mathbf W^{1/2}$ and the shared operators
  $\mathbf C\mathbf Q\mathbf K\mathbf D\mathbf P$ gives the weighted
  model-to-data Jacobian $\mathbf J_d$.  For coincident grids,
  $\mathbf R=\mathbf I$.}
  \label{fig:implicit_factorized_concept}
\end{figure}

For an individual observation $i$, set $\Psi_g=d_i$.  Receiver sampling
supplies the dependence on the converged field, while $T_0$ supplies the
explicit source-slowness contribution.  Let
$\widetilde{\boldsymbol{\tau}}_g$ denote the full nodal factored field,
including its fixed source-neighborhood values, and let
$\widetilde{\mathbf{r}}_i$ be the sparse full-grid vector containing the four
bilinear receiver weights.  The discrete implementation realizes
$\mathcal{R}_i$ by interpolating the factored field and evaluating $T_0$ at
the receiver:
\begin{equation}
  d_i=T_{0,g}(\mathbf{x}_{r,i})
      \widetilde{\mathbf{r}}_i^T
      \widetilde{\boldsymbol{\tau}}_g,
  \label{eq:t0_tau_receiver}
\end{equation}
Let $\mathbf{r}_i$ denote the restriction of
$\widetilde{\mathbf{r}}_i$ to the free-node space of
$\boldsymbol{\tau}_g$; fixed-node values contribute to $d_i$ but not to its
derivative with respect to the free field.  The receiver derivatives are
\begin{equation}
  \mathbf{h}_i
  =\nabla_{\boldsymbol{\tau}_g}d_i
  =T_{0,g}(\mathbf{x}_{r,i})\mathbf{r}_i,
  \qquad
  \mathbf{d}_i^{\mathrm{src}}
  =\left.\nabla_{\mathbf{s}_{p_g,f}}d_i
   \right|_{\boldsymbol{\tau}_g}
  =d_{\mathcal{S}}(\mathbf{x}_{r,i},\mathbf{x}_{s,g})
   \left(\widetilde{\mathbf{r}}_i^T
         \widetilde{\boldsymbol{\tau}}_g\right)
   \boldsymbol{\chi}_{s,g}.
  \label{eq:receiver_derivatives}
\end{equation}
Here $\mathbf h_i$ measures the sensitivity of the sampled traveltime to
the converged free-node field, whereas $\mathbf d_i^{\mathrm{src}}$ is the
explicit source-factor contribution at fixed $\boldsymbol\tau_g$.  The
corresponding adjoint solve from
equation~\eqref{eq:general_discrete_adjoint} is
\begin{equation}
  \mathbf{B}_g^T\boldsymbol{\lambda}_i=\mathbf{h}_i
  \label{eq:individual_discrete_adjoint}
\end{equation}
and equation~\eqref{eq:general_discrete_gradient} gives the sensitivity on
the phase-velocity grid by the chain rule:
\begin{equation}
  \mathbf{a}_i
  =\nabla_{\mathbf{s}_{p_g}}d_i
  =\mathbf{R}^T\left(
    \mathbf{d}_i^{\mathrm{src}}
    +\mathbf{U}_{s,g}^T\boldsymbol{\lambda}_i\right).
  \label{eq:individual_row}
\end{equation}
Here $\mathbf R^T$ returns the sensitivity to the phase-velocity grid using
the forward interpolation weights.  Thus $\mathbf a_i$ is the
phase-slowness derivative of observation $i$ retained on this grid, and
$\mathbf a_i^T$ forms one sparse row of the gather Jacobian $\mathbf A_g$
in equation~\eqref{eq:gather_A} and hence of $\mathbf A$.
Figure~\ref{fig:implicit_factorized_concept} summarizes this construction:
panel (a) shows the derivation of $\mathbf a_i$ for one receiver,
and panel (b) places the returned observation rows in the full factorized
model-to-data Jacobian.

The same adjoint equation also gives the gather data-misfit gradient
without forming the individual rows.  For gather $g$, define its
contribution to the weighted objective in equation~\eqref{eq:data_objective}
as $\Phi_{d,g}=\frac12\sum_{i\in\mathcal I_g}w_i e_i^2$, and let
$q_i=w_i e_i$.
One aggregate adjoint solve uses the weighted receiver terms as its
right-hand side:
\[
  \mathbf B_g^T\boldsymbol\lambda_{\Phi,g}
  =\sum_{i\in\mathcal I_g}q_i\mathbf h_i.
\]
By linearity, this solution is the $q_i$-weighted sum of the
individual adjoints in
equation~\eqref{eq:individual_discrete_adjoint}.
The direct source derivative and transpose interpolation obey the same
linear combination, giving
\begin{equation}
  \nabla_{\mathbf s_{p_g}}\Phi_{d,g}
  =\sum_{i\in\mathcal I_g}q_i\mathbf a_i.
  \label{eq:aggregate_individual_equivalence}
\end{equation}
Thus, the aggregate adjoint gives exactly the weighted sum of the
individual traveltime sensitivity kernels.

\subsubsection{Graph-based evaluation of individual kernels}
\label{sec:graph_kernel_evaluation}

Equation~\eqref{eq:individual_row} reduces construction of one row of
$\mathbf A$ to the receiver-dependent adjoint solve in
equation~\eqref{eq:individual_discrete_adjoint}.  To evaluate many such
rows efficiently, we exploit the sparse active dependency graph of the
converged Eikonal system.

Bilinear receiver sampling confines $\mathbf h_i$ to at most four
forward-grid nodes.  For the two-dimensional surface-propagation problem
considered here, each row of $\mathbf B_g$ contains a unit diagonal and at
most two off-diagonal coefficients, because the active first-order stencil
depends on at most one upwind neighbor in each surface coordinate.  These
dependencies define a sparse directed graph.
In the transpose solve, contributions from $\mathbf{h}_i$ follow this
graph toward the source.  The nodes reached from the receiver form its
ancestor closure $\mathcal{V}_{a,i}$.  For a fixed dependency graph,
$\boldsymbol{\lambda}_i$ vanishes outside this closure, so the adjoint
equation reduces exactly to
\begin{equation}
  \mathbf{B}_{g,\mathcal{V}_{a,i}\mathcal{V}_{a,i}}^{T}
  \boldsymbol{\lambda}_{i,\mathcal{V}_{a,i}}
  =\mathbf{h}_{i,\mathcal{V}_{a,i}}.
  \label{eq:restricted_individual_adjoint}
\end{equation}
Fast sweeping does not impose a global acceptance order, so the selected
dependencies can contain cycles.  In a two-node cycle
$j\leftrightarrow k$, for example, the adjoint values at $j$ and $k$
must be found together.  We group mutually dependent nodes into strongly
connected components (SCCs).  Collapsing each component to one block
gives an acyclic graph and a block-triangular ordering of $\mathbf{B}_g^T$.
For each receiver, we visit only the blocks in its ancestor closure,
processing acyclic single-node blocks by scalar accumulation and cyclic
blocks by coupled solves.

The graph and its decomposition are built once per gather and reused for
all receiver solves.  For an acyclic closure, the
receiver solve costs $O(N_{a,i}+E_{a,i})$, where $N_{a,i}$ and $E_{a,i}$
count its nodes and active edges.  Coupled block solves are described in
\ref{app:cycles}.

\section{Results}
\label{sec:results}

We evaluate the formulation through derivative tests, kernel and cost
comparisons, synthetic inversions, and an application to the contiguous
United States.  Baseline comparisons with SurfATT use coincident phase-velocity
and forward grids ($r=1$, $\mathbf{R}=\mathbf{I}$) to retain its forward
discretization.  Forward-grid refinement is evaluated in the derivative and
kernel tests and in an additional synthetic inversion with $r=2$.
The U.S. application also uses $r=2$.

\subsection{Consistency and Efficiency of Individual Traveltime Kernels}
\label{sec:kernel_results}

We verified the complete weighted operator chain
$\mathbf{J}_d=\mathbf{W}^{1/2}\mathbf{A}\mathbf{C}\mathbf{Q}
\mathbf{K}\mathbf{D}\mathbf{P}$ using deterministic model- and data-space
directions in a smooth, asymmetric three-dimensional model.  The $V_S$ field
on its $9\times8\times21$ physical grid had a background increasing with
depth and weak sinusoidal lateral variations.  The nonlinear test included
control interpolation, layered dispersion at multiple periods,
off-grid source and receiver sampling, and discrete Eikonal propagation.  For
a control-space direction $\mathbf{v}$, define the weighted prediction
$\mathbf{d}_w(\boldsymbol{\beta})=\mathbf{W}^{1/2}\mathbf{d}(\boldsymbol{\beta})$, with
$\mathbf{W}$ fixed during the test.  Perturbed models are
$\boldsymbol{\beta}_{\epsilon}=\boldsymbol{\beta}\odot
\exp(\epsilon\mathbf{P}\mathbf{v})$.  The normalized remainder
\begin{equation}
  E(\epsilon)=
  \frac{\|\mathbf{d}_w(\boldsymbol{\beta}_{\epsilon})
  -\mathbf{d}_w(\boldsymbol{\beta})-\epsilon\mathbf{J}_d\mathbf{v}\|_2}
  {\max(\|\mathbf{d}_w(\boldsymbol{\beta})\|_2,1)}
  \label{eq:full_chain_taylor}
\end{equation}
decreased at second order, with a mean observed order of 2.02
(Table~\ref{tab:full_chain_validation}).  The complementary transpose test
gave a relative difference of $1.32\times10^{-16}$ between
$\langle\mathbf{J}_d\mathbf{v},\mathbf{q}\rangle$ and
$\langle\mathbf{v},\mathbf{J}_d^T\mathbf{q}\rangle$.
We repeated this test for forward-grid refinement factors $r=1,2,4,8$,
holding the original model, observations, weights, and perturbation
direction fixed.  Both factored
$T_0\tau$ receiver sampling and interpolation of nodal total time gave mean
Taylor orders of 2.02--2.04 and transpose errors below
$1.89\times10^{-16}$ (Table~\ref{tab:full_chain_validation}).
The inversions use the $T_0\tau$ operator in
equation~\eqref{eq:t0_tau_receiver}.

\begin{table}[t]
\centering
\caption{Taylor and transpose tests of the complete weighted model-to-data
derivative.  The upper block gives the $r=1$, $T_0\tau$ test;
the lower block summarizes the same test at four refinement factors for
both receiver operators.  Mean orders use the four perturbation amplitudes
listed in the upper block.}
\label{tab:full_chain_validation}
\begin{tabular}{lccc}
\hline
Test & $\epsilon$ & Relative error & Observed order \\
\hline
Taylor & $1.6\times10^{-2}$ & $3.085\times10^{-5}$ & 2.000 \\
Taylor & $8.0\times10^{-3}$ & $7.712\times10^{-6}$ & 2.019 \\
Taylor & $4.0\times10^{-3}$ & $1.903\times10^{-6}$ & 2.054 \\
Taylor & $2.0\times10^{-3}$ & $4.584\times10^{-7}$ & --- \\
Transpose & --- & $1.318\times10^{-16}$ & --- \\
\hline
\end{tabular}
\par\medskip
\begin{tabular}{ccccc}
\hline
& \multicolumn{2}{c}{$T_0\tau$ sampling}
& \multicolumn{2}{c}{Total-time sampling} \\
$r$ & Mean order & Transpose error & Mean order & Transpose error \\
\hline
1 & 2.024 & $1.318\times10^{-16}$ & 2.024 & $6.583\times10^{-17}$ \\
2 & 2.036 & $6.371\times10^{-17}$ & 2.036 & $6.371\times10^{-17}$ \\
4 & 2.038 & $1.883\times10^{-16}$ & 2.038 & $6.279\times10^{-17}$ \\
8 & 2.040 & $6.217\times10^{-17}$ & 2.040 & $6.217\times10^{-17}$ \\
\hline
\end{tabular}
\end{table}

\begin{figure}[t]
  \centering
  \includegraphics[width=0.96\linewidth]{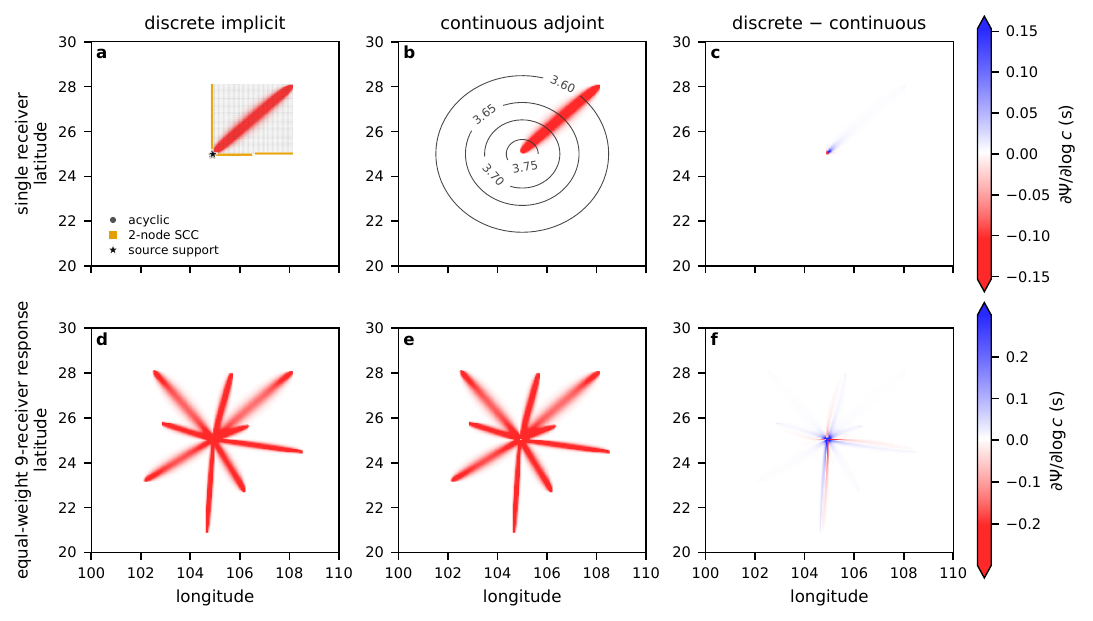}
  \caption{Representative $\log c$ kernels in the $200\times200$ synthetic
  model.  (a--c) Kernel for one representative receiver; (d--f) kernel for an
  equally weighted response of all nine receiver traveltimes.  The columns
  show the discrete implicit calculation, the continuous-adjoint calculation,
  and their
  difference.  Symbols overlaid in (a) show the exact active dependency set:
  gray points denote acyclic ancestor nodes, orange squares denote nodes in
  two-node strongly connected components, and stars denote source-support nodes.
  Contours in (b) show phase velocity in km~s$^{-1}$.  The plotted quantities
  are $\partial\Psi/\partial\log c$, with $\Psi=d_i$ in the upper row and
  $\Psi=\sum_{i=1}^{9}d_i$ in the lower row.  A common symmetric color scale
  within each row is set by the 95th percentile of the nonzero absolute
  discrete-kernel coefficients; larger magnitudes saturate at the color
  limits.}
  \label{fig:kernel_maps}
\end{figure}

We then examined traveltime sensitivity kernels in a separate two-dimensional
phase-velocity model with a central Gaussian high-velocity anomaly.  The
field was sampled on a $200\times200$ surface grid
(Figure~\ref{fig:kernel_maps}b).  One source near the anomaly center was paired
with nine off-grid receivers.

For each receiver, we computed an implicit-differentiation kernel and a kernel
using the continuous-adjoint formulation of SurfATT \citep{HaoEtAl2024}; both
were expressed as traveltime derivatives with respect to $\log c$,
including the sign from $\delta s=-s\,\delta\log c$.
Figure~\ref{fig:kernel_maps} compares
the resulting kernels for one representative receiver and for an equally
weighted response of all nine receiver traveltimes.  The active dependency
set overlaid in
Figure~\ref{fig:kernel_maps}a shows that most selected nodes are acyclic and
that every cyclic SCC contains only two nodes.  The implicit and continuous
calculations produce similar dominant source--receiver sensitivity patterns
for both responses.  The remaining differences reflect their respective
numerical representations of the source, Eikonal operator, and receiver
sampling.

\begin{figure}[t]
  \centering
  \includegraphics[width=0.96\linewidth]{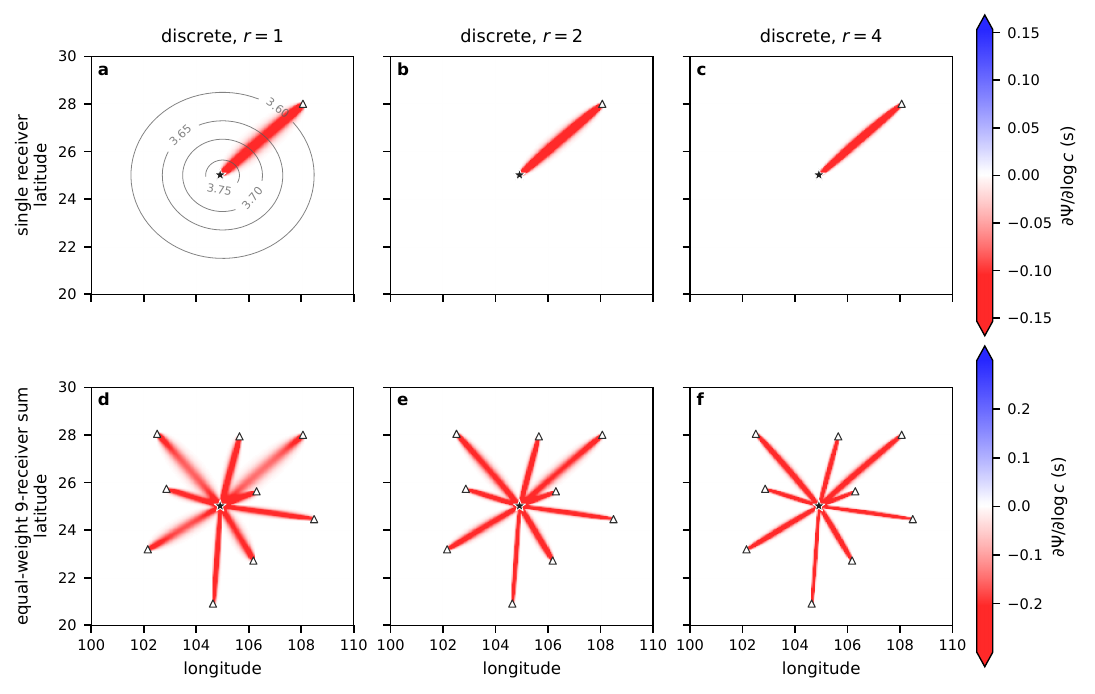}
  \caption{Effect of Eikonal refinement on discrete $\log c$ kernels for
  the model and observations in Figure~\ref{fig:kernel_maps}.
  (a--c) The same representative receiver; (d--f) the equally weighted sum
  of all nine receiver traveltimes.  Columns show $r=1$, $r=2$, and $r=4$.
  All kernels refer to the same $200\times200$ phase-velocity grid after
  transpose interpolation and conversion to $\log c$ derivatives.
  Stars mark the source and triangles mark receivers;
  contours in (a) show phase velocity in km~s$^{-1}$.  Each row shares the
  symmetric color limits set by the 95th percentile of nonzero absolute
  $r=1$ coefficients, with larger magnitudes saturated.}
  \label{fig:kernel_refinement}
\end{figure}

We next compared the discrete kernels at $r=1$, $2$, and $4$, keeping the same
$200\times200$ phase-velocity model, source, and receivers
(Figure~\ref{fig:kernel_refinement}).  Bilinear slowness interpolation
defined forward grids of $200\times200$, $399\times399$, and $797\times797$
nodes, respectively.  All calculations used the $T_0\tau$ receiver operator,
with kernels returned by $\mathbf{R}^T$ and converted to $\log c$
derivatives on the phase-velocity grid.  As the forward grid becomes finer,
the dominant kernel weights concentrate in narrower bands around the
source--receiver ray paths, approaching the localized sensitivity assumed
in ray theory.

\begin{figure}[t]
  \centering
  \includegraphics[width=\linewidth]{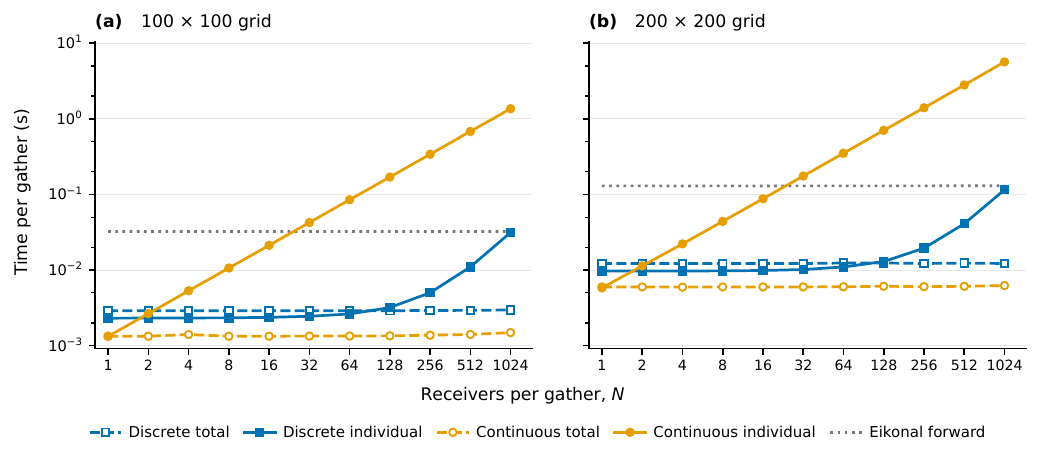}
  \caption{Time per gather versus receiver count on the 100-by-100 (a) and
  200-by-200 (b) grids.  Blue and orange denote discrete and continuous
  calculations.  Dashed lines (total) show one aggregate calculation; solid
  lines (individual) show the cumulative cost of $N$ single-receiver
  calculations.  Forward time is shown separately in gray.
  Discrete curves include graph preparation and kernel construction, but
  discrete total excludes receiver-load assembly.  Continuous curves include
  load assembly, coefficient preparation, initialization, and field solves;
  preparation is repeated for individual calls.  Continuous field-to-kernel
  conversion is excluded.  Timing components are listed in
  \ref{app:propagation_timing}.
  Measurements used one CPU core, five warmups, and 15 repetitions.
  Lines and shading show medians and 5th--95th percentiles of stage sums
  per repetition.}
  \label{fig:receiver_scaling}
\end{figure}

We compared discrete and continuous calculations using the same model and
source on $100\times100$ and $200\times200$ grids, with 1--1024 off-grid
receivers (Figure~\ref{fig:receiver_scaling}).  For each method, \emph{total}
denotes one aggregate solve with all receiver contributions combined in its
right-hand side; discrete total uses the SCC-condensed aggregate adjoint.
\emph{Individual} denotes the cumulative cost of $N$ separate single-receiver
calculations.  All four workloads share one converged forward field.
The timings include method-specific preparation and adjoint calculations;
the common Eikonal forward cost is shown separately.  The figure caption
specifies the timed operations, and \ref{app:propagation_timing} lists
their measured components.

The aggregate costs remained nearly independent of receiver count, whereas
the cost of separate receiver responses increased with $N$.  Repeated
continuous calls, each including coefficient preparation, scaled approximately
linearly; shared graph preparation kept discrete individual costs nearly flat
for small gathers.  Work for each receiver dominated as gathers grew.

At 1024 receivers, discrete individual construction took 31.5 and 115.0~ms
on the $100\times100$ and $200\times200$ grids, respectively, compared with
1.36 and 5.65~s for repeated continuous solves.  The corresponding measured
speedups were 43.2 and 49.1, respectively.

Within the discrete implementation, constructing all 1024 kernels cost
10.6 and 9.4 times one aggregate response, which took 2.96 and 12.20~ms.
Including the common forward solves of 32.3 and 130.0~ms, the corresponding
cost ratios decreased to 1.81 and 1.72.  Thus, computing the forward field
and all 1024 individual kernels took less than twice the time required for
the forward field and one aggregate response.

\subsection{Scaling of the Factorized Jacobian}
\label{sec:factorized_scaling_results}

\begin{figure}[t]
  \centering
  \includegraphics[width=\linewidth]{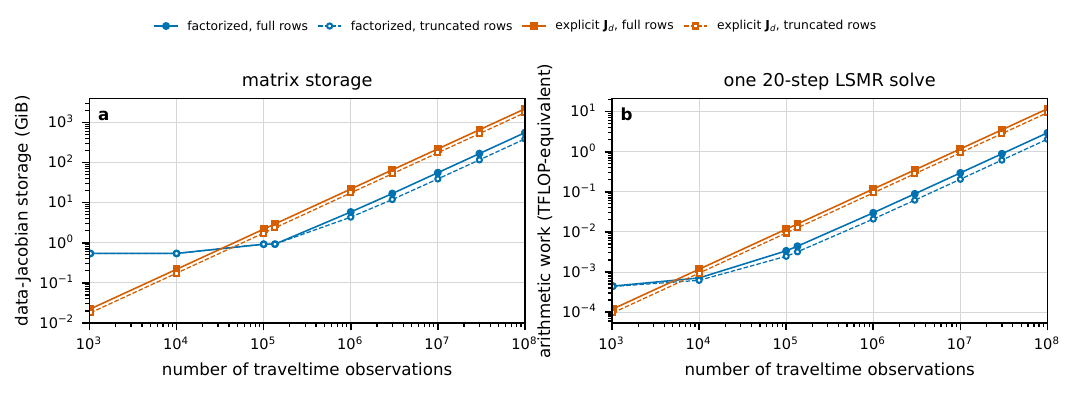}
  \caption{Estimated scaling of (a) data-Jacobian storage and (b) arithmetic
  work for a 20-step LSMR solve as a function of the number of traveltime
  observations.  The estimates are calibrated using 134,820 observations from
  the checkerboard experiment, a $0.2^{\circ}$ forward grid, and 7,128
  controls.  Both representations use 64-bit values and 64-bit CSR column
  indices in the storage estimate, with the factorized allocation and
  workspace overheads retained.  The grid, period set, and calibrated mean
  row supports are held fixed as the observation count increases.
  Factorized curves retain the shared operators, whereas explicit
  $\mathbf{J}_d$ denotes the explicitly composed weighted data Jacobian; its support is
  estimated from 1,024 sampled rows.  Solid curves retain all coefficients in
  $\mathbf{A}$, and dashed curves apply a relative cutoff of $10^{-3}$.  Work
  includes one Jacobian and one transpose-Jacobian product per LSMR step.}
  \label{fig:factorized_scaling}
\end{figure}

Figure~\ref{fig:factorized_scaling} compares Jacobian storage and LSMR work for
the factorized and explicitly composed representations, calibrated using the
synthetic checkerboard data set described in
Section~\ref{sec:small_anomaly_results}.  Both CSR representations are
accounted for with 64-bit values and 64-bit column indices.  At small data
volumes, storage of the shared factors is comparable to that of the
observation-dependent rows.  As the number of observations increases, the
difference in stored coefficients per row increasingly determines storage
and matrix-vector work.  For the 134,820 traveltimes used in the inversion, the untruncated
factorized representation reduced the estimated storage and LSMR arithmetic
by factors of 3.19 and 3.59, respectively, relative to the explicit matrix.
Extrapolating the same grid, period set, and mean row supports to $10^8$
traveltimes gives reductions of 3.93 in storage and 3.97 in arithmetic.

These large-data gains reflect the difference in mean row supports.  The
untruncated $\mathbf A$ has a mean of 368.7 coefficients per row, whereas
propagation of 1,024 sampled row supports through the shared mappings gives
approximately 1,464.0 control-grid coefficients per composite row.  The
ratio is 3.97, greater than $N_c/N_h^{\phi}=7128/2601\simeq2.74$ because
interpolation expands the horizontal support on the control grid, as
discussed in Section~\ref{sec:storage_scaling}.

We also tested storage compression by truncating small coefficients in each
row of $\mathbf A$.  The four source interpolation nodes and nodes within
two grid intervals of the source were always retained.  Elsewhere,
coefficients with magnitudes below $10^{-3}$ of the row's maximum absolute
coefficient outside this protected region were omitted.  This rule removed
31\% of the coefficients, with a relative Frobenius error of
\[
  \frac{\|\mathbf{A}_{\mathrm{trunc}}-\mathbf{A}\|_F}{\|\mathbf{A}\|_F}
  =5.3\times10^{-4}.
\]
The dashed curves in Figure~\ref{fig:factorized_scaling} show the associated
large-data scaling.

\subsection{Recovery of Synthetic Anomaly Patterns}
\label{sec:small_anomaly_results}

We constructed a three-dimensional checkerboard of alternating positive and
negative $\log V_S$ anomalies with a peak absolute log perturbation of 0.05
relative to a horizontally uniform background.  Horizontal lobes span
approximately $1.8^{\circ}$, and two vertical lobes are defined in normalized
background log velocity.  The background, which also served as the initial
model, increases from 3.0~km~s$^{-1}$ at the surface to 4.35~km~s$^{-1}$ at
100~km depth.  The target log-velocity anomaly was constructed in the adopted
control space, and
synthetic traveltimes received 0.5\% relative Gaussian noise.  The full
background profile and target construction are specified in
\ref{app:optimizer_parameters}.  The common synthetic domain used a $51\times51\times51$
physical grid with $0.2^{\circ}$ horizontal and 2~km vertical spacing.
The $r=1$ comparisons used coincident $51\times51$ phase-velocity and
forward grids for synthesis and inversion, matching the original SurfATT discretization.
An $18\times18\times22$ control grid contained 7,128 $\log V_S$ parameters
at nominal spacings of $0.6^{\circ}$ and 5~km.  The acquisition comprised 123 stations,
30 periods between 5 and 50~s, and 134,820 interstation paths.  All inversions
ran on one AMD EPYC 7453 node using 24 single-threaded MPI ranks.  For this
experiment, $V_P$ and density were derived from $V_S$ using crustal
empirical relations \citep{Brocher2005}.

At $r=1$, we compared Gauss--Newton (GN), L-BFGS, and gradient
descent (GD) using the same discrete derivative, $T_0\tau$ receiver
interpolation, and control-space log-velocity parameterization.
L-BFGS and GD obtained their data gradients from an aggregate discrete
adjoint, while GN retained individual Jacobian rows.  All methods used
exponential updates bounded by $\max|\Delta\log V_S|\leq0.02$ on the physical grid.
All observations had unit weight ($\mathbf{W}=\mathbf{I}$).
The GN runs used one control grid and solved the regularized problem
in equation~\eqref{eq:linearized_inverse} using at most 20 LSMR iterations
per nonlinear update.  Their
horizontal and vertical reference roughness levels were 5\% and 20\% of the
corresponding diagonal-curvature scales defined in
equation~\eqref{eq:regularization_scaling}, and the diagonal step-damping
ratio was held at 0.5\%.  The L-BFGS runs applied cumulative
second-order roughness to the current absolute $V_S$ model and used no
damping.  The formal GD run used neither roughness nor damping.
The comparison evaluates the combined effects of the update method,
regularization, and step acceptance.

The eight $r=1$ runs shared the observations, noise, grids, and
initial model, giving the same initial RMS of 1.199196~s.
An additional GN run used the reference parameters with $r=2$ on a
$101\times101$ forward grid, retaining the same phase-velocity and control
grids, target, and initial model.  Its observations were regenerated on this
refined grid using the same relative-noise realization, giving an initial
RMS of 1.280767~s.

\begin{figure}[t]
  \centering
  \includegraphics[width=\linewidth]{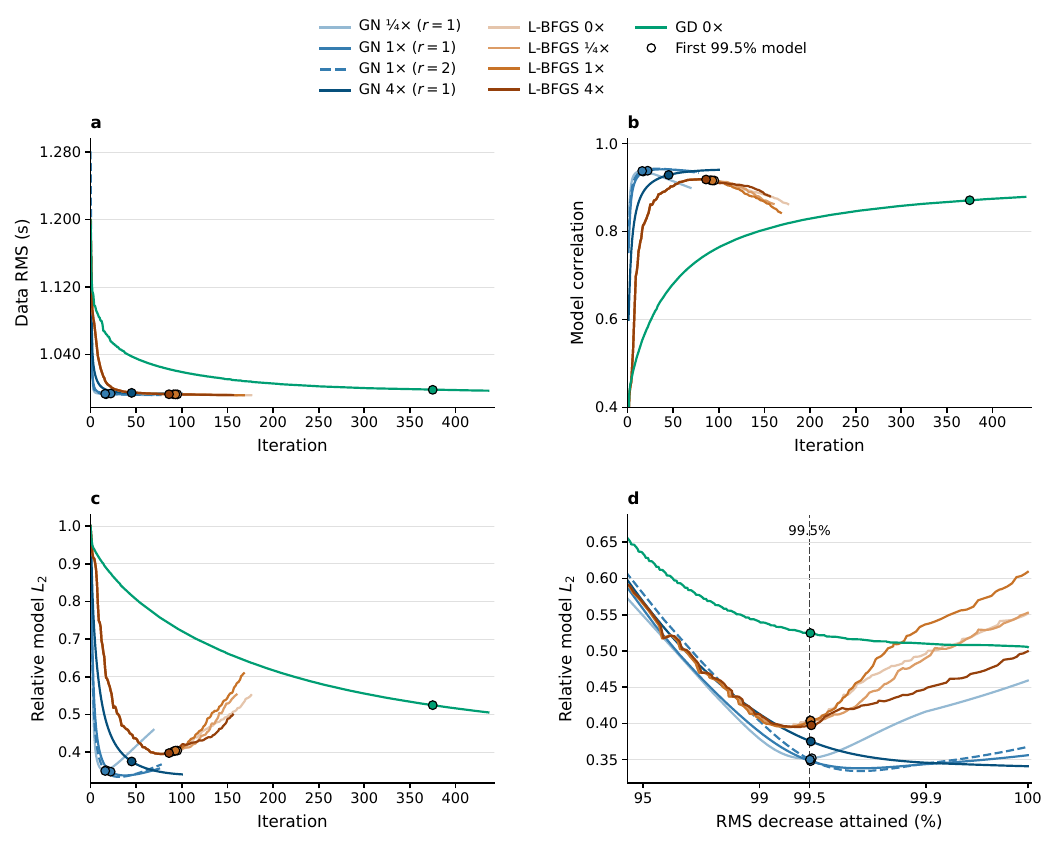}
  \caption{Convergence of the nine formal checkerboard inversions.  Panels show
  (a) traveltime RMS, (b) anomaly correlation, and (c) relative anomaly $L_2$
  error as functions of iteration, and (d) the same $L_2$ error as a
  function of the fraction of RMS decrease attained over each complete run.
  Circles mark the first saved model satisfying the 99.5\% criterion;
  the vertical dashed line in (d) marks this threshold.
  Solid curves use $r=1$; the dashed blue curve is the reference GN run at
  $r=2$ with observations synthesized at $r=2$.  All cases share the target,
  initial model, acquisition, and relative-noise realization.
  Roughness multipliers refer to each method's reference horizontal and
  vertical strengths; the GN and L-BFGS terms act on the update
  and current absolute model, respectively.}
  \label{fig:small_regularization}
\end{figure}

Reference regularization strengths were selected based on model recovery
in the parameter searches (\ref{app:optimizer_parameters}).  The formal comparison
at $r=1$ used GN roughness multipliers of $0.25$, $1$, and $4$, L-BFGS multipliers
of $0$, $0.25$, $1$, and $4$, and unregularized GD.  The L-BFGS reference
weights were $2\times10^{-6}$ horizontally and $1\times10^{-6}$ vertically.
Figure~\ref{fig:small_regularization}
shows all nine formal runs.

Let $R_k$ denote the traveltime RMS after update $k$.  Each formal run allowed
at most 2,000 updates and terminated after five successive updates with
insufficient RMS improvement,
$(R_{k-1}-R_k)/R_{k-1}<10^{-5}$ (0.001\%); an RMS increase also counts as
insufficient improvement.  All nine runs ended by this RMS rule.

Model recovery is measured on the complete $51\times51\times51$ physical
grid using anomalies relative to the common initial background:
\begin{equation}
  \mathbf{u}_k=\log(\boldsymbol{\beta}_k/\boldsymbol{\beta}_0),
  \qquad
  \mathbf{u}_*=\log(\boldsymbol{\beta}_*/\boldsymbol{\beta}_0),
  \qquad
  E_{m,k}=\frac{\|\mathbf{u}_k-\mathbf{u}_*\|_2}{\|\mathbf{u}_*\|_2}.
  \label{eq:synthetic_recovery_metrics}
\end{equation}
Here $\boldsymbol{\beta}_*$ is the synthetic target velocity model, and
$\mathbf{u}_*$ is its log-velocity anomaly relative to the initial model
$\boldsymbol{\beta}_0$; $\mathbf{u}_k$ is the recovered anomaly after $k$
updates.  The logarithms and divisions are component-wise.  We report the
relative model $L_2$ error $E_{m,k}$ and the linear correlation coefficient
between $\mathbf{u}_k$ and $\mathbf{u}_*$.  Both metrics use equal weight
over the full physical grid and the original anomaly amplitudes.

Over the complete runs, the $r=1$ GN configurations stopped after 69--100
updates, L-BFGS after 156--176, and GD after 436, with terminal RMS values
between 0.9916 and 0.9971~s.  Several histories show late-stage overfitting,
with model error increasing during the final small reductions in RMS,
especially for L-BFGS and the weakest GN roughness.

\begin{figure}[t]
  \centering
  \includegraphics[width=\linewidth]{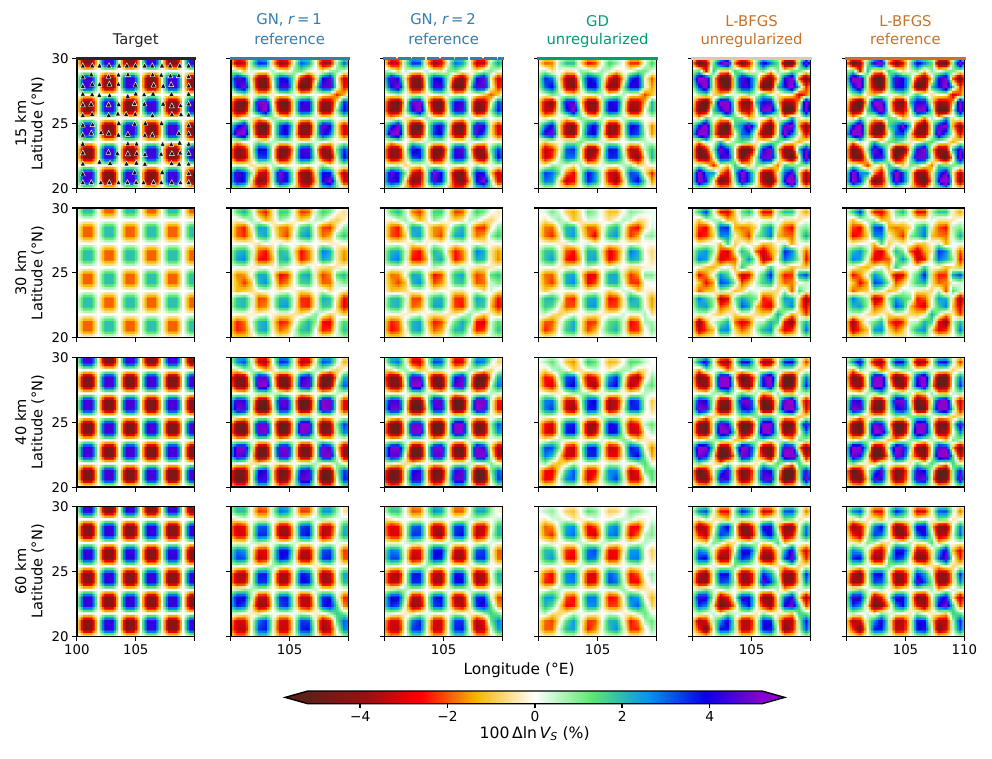}
  \caption{Horizontal slices through the checkerboard target and five selected
  inversion results at 15, 30, 40, and 60~km depth.  Columns show the target,
  reference GN results at $r=1$ and $r=2$ in adjacent columns,
  unregularized GD, and L-BFGS with zero and reference cumulative roughness.
  Data synthesis and inversion use the same $r$ in each case.  Each result is the
  first saved model satisfying the 99.5\% criterion.
  Triangles in the upper-left target panel show station locations.  All panels
  show $100\,\Delta\ln V_S=100\log(V_S/V_{S,0})$ relative to the common
  initial background on a common scale.}
  \label{fig:small_horizontal}
\end{figure}

\begin{figure}[t]
  \centering
  \includegraphics[width=\linewidth]{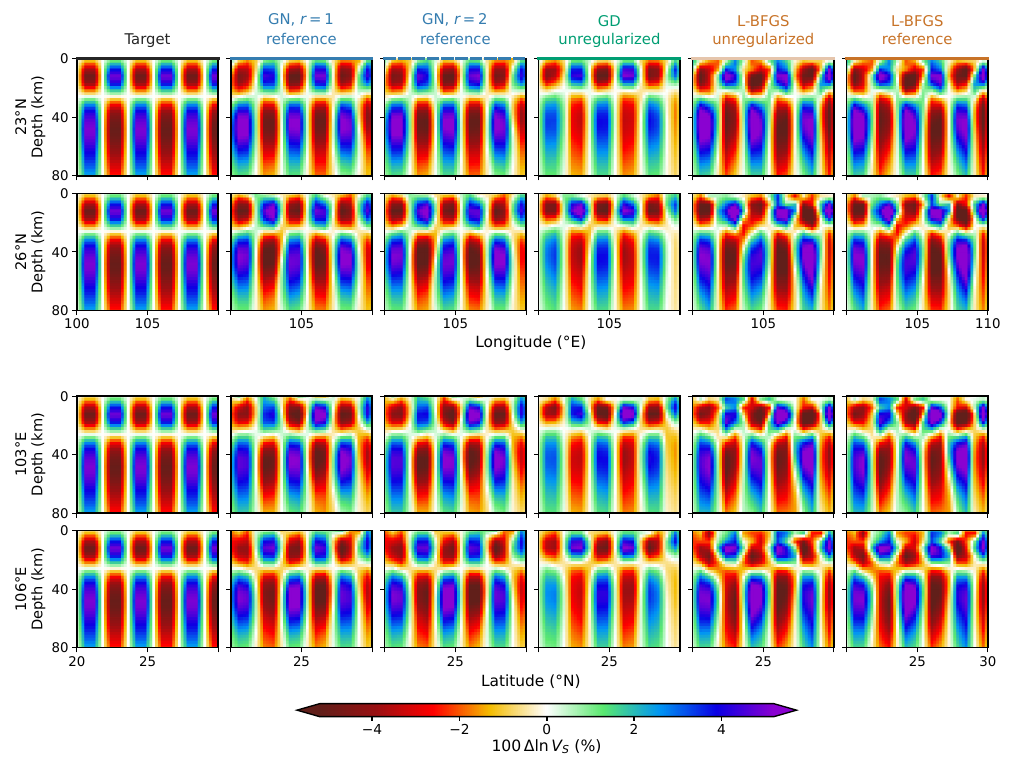}
  \caption{Four vertical sections through the checkerboard target and the five
  selected inversion results displayed in Figure~\ref{fig:small_horizontal}.
  Section locations are labeled in the figure, and the common scale shows
  $100\,\Delta\ln V_S=100\log(V_S/V_{S,0})$ relative to the common initial
  background.}
  \label{fig:small_vertical}
\end{figure}

To compare models at a common stage of data fitting, we define the fraction
of the total RMS decrease attained in each run:
\begin{equation}
  \xi_k = \frac{R_0-R_k}{R_0-\min_j R_j},
  \label{eq:rms_decrease_attained}
\end{equation}
where $\min_j R_j$ is the lowest RMS in that run.
Figures~\ref{fig:small_horizontal} and~\ref{fig:small_vertical} use the first
saved model with $\xi_k\geq0.995$.

The reference GN run at $r=1$ first satisfied this criterion after 22 updates and
2.6~min.  Reference and unregularized L-BFGS required 93 and 95 updates,
respectively, and about 6.5 and 6.4~min; GD required 375 updates and
76.4~min.
The corresponding model errors were 0.348 for GN, 0.404 for both L-BFGS
models, and 0.525 for GD, at RMS values between 0.9927 and 0.9981~s.

Figures~\ref{fig:small_horizontal} and~\ref{fig:small_vertical} compare the
selected models on a common color scale.  In these tests, GN recovers the alternating
anomaly pattern and its depth variation more faithfully than L-BFGS and GD,
consistent with the lower relative model errors reported above.  The
reference and unregularized L-BFGS results are similar to each other,
while GD shows weaker recovery toward the sparsely sampled margins.
The $r=2$ GN run first satisfied the criterion at update 16 in 4.8~min, with RMS
0.9934~s and model error 0.350, close to the $r=1$ reference value of 0.348.
Forward-grid refinement narrowed the discrete kernels, while model recovery
remained comparable in the matched-grid synthetic experiments.
The minimum relative model errors were 0.338 at update 37 for $r=1$ and
0.334 at update 31 for $r=2$.

\subsection{Application Across the Contiguous United States}
\label{sec:continental_results}

\begin{figure}[t]
  \centering
  \includegraphics[width=\linewidth]{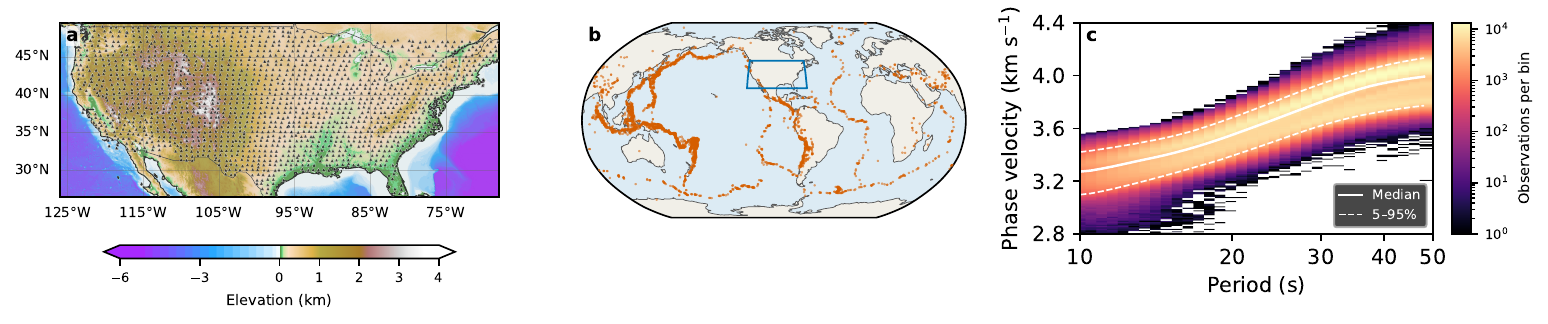}
  \caption{Data distribution for the contiguous-US application.  (a) Station
  locations retained after quality control, over topography.  (b) Locations
  of the teleseismic earthquakes
  used in the two-station measurements; the blue box marks the study region.
  (c) Phase-velocity distribution of the retained observations over period.
  White solid and dashed curves show the median and 5th--95th percentiles,
  respectively, and color denotes observation counts in
  0.01~km~s$^{-1}$ bins at each measured period.}
  \label{fig:usa_data}
\end{figure}

Broadband waveforms recorded at 1,668 stations across the contiguous United
States were accessed through the IRIS Data Management Center
\citep{TrabantEtAl2012}.  The network was dominated by the USArray
Transportable Array and supplemented by
permanent and regional stations (Figure~\ref{fig:usa_data}a).  Interstation
funda\-mental-mode Rayleigh-wave phase velocities were extracted from
teleseismic records of 3,857 earthquakes between 2004 and 2019 using the
two-station method \citep{MeierEtAl2004}, implemented with the multiple-filter
technique \citep{DziewonskiHales1972}.  Quality control applied criteria for
interstation distance, phase-velocity uncertainty, and the minimum number
of accepted periods per pair, followed by screening of preliminary-inversion
residuals.  The final data set contains 6,340,798 interstation phase
traveltimes at 33 periods
between 10 and 50~s, sampling 1,662 stations.
Figure~\ref{fig:usa_data}b--c shows the event distribution and the
period-dependent distribution of the retained phase velocities.

For path length $L_i$, phase velocity $c_i$, and phase-velocity uncertainty
$\sigma_{c,i}$ (with a 0.005~km~s$^{-1}$ floor), the base traveltime
uncertainty was
$\sigma_{t,i}=L_i\sigma_{c,i}/c_i^2$.  The raw inverse-variance weights
$w_i^{\mathrm{raw}}=\sigma_{t,i}^{-2}$ were normalized by their median at
each period and clipped to the range 0.1--10 to obtain
$w_i^{\mathrm{QC}}$.  These weights were held
fixed throughout the inversion, with $[\mathbf{W}]_{ii}=w_i^{\mathrm{QC}}$.

The initial three-dimensional $V_S$ model was derived from the US.2016
$V_{SV}$ model \citep{ShenRitzwoller2016} by filling uncovered cells,
smoothing the completed volume, and enforcing a monotonic depth profile
before extension to 120~km.  Processing details and the resulting initial
model are given in \ref{app:usa_initial_model}.

The physical model used $127\times59\times61$ nodes, with $0.5^{\circ}$
horizontal spacing and 2~km depth spacing to 120~km.  Phase velocities were
calculated on its $127\times59$ horizontal grid.  Bilinear slowness
interpolation with $r=2$ defined a $253\times117$ forward grid at
$0.25^{\circ}$ spacing.  For each source--period gather, the Eikonal solve
used an adaptively guarded window defined by the source and receiver
distribution.  The three-dimensional model was parameterized using
$n_{\mathrm{comp}}=5$ equally weighted, staggered
$65\times31\times18$ control grids
\citep{TongEtAl2019MultipleGrid},
corresponding to 36,270 coefficients per grid (181,350 in total), with nominal
spacings of $1^{\circ}$ horizontally and 7.5~km vertically.  For this application and its
checkerboard test, the crustal empirical relations \citep{Brocher2005} were
used where $V_S\leq4.0$~km~s$^{-1}$, whereas the mantle branch
$V_P=V_{P,\mathrm M}(z)[V_S/V_{S,\mathrm M}(z)]^{0.5}$ and
$\rho=\rho_{\mathrm M}(z)[V_S/V_{S,\mathrm M}(z)]^{0.3}$ was used where
$V_S\geq4.5$~km~s$^{-1}$.  The prescribed reference profiles are defined in
Table~\ref{tab:usa_mantle_reference}.  Between the two bounds, the branches
were blended with mantle weight $3t^2-2t^3$, where $t=(V_S-4.0)/0.5$ with
$V_S$ in km~s$^{-1}$.

The inversion used 52 single-threaded MPI ranks across two CPU nodes:
28 ranks on an AMD EPYC 7453
processor with 28 physical cores and 24 ranks on a dual-socket Intel Xeon Gold
6226 system with 24 physical cores.  Each control grid used
at most 20 LSMR iterations per nonlinear update.  The five physical-grid
log-velocity increments were averaged before applying a common step bound
and line search, as in equation~\eqref{eq:component_log_average}.
The GN line search used the same data-misfit-based trial selection described
in \ref{app:optimizer_parameters}.  Both the real-data
inversion and its checkerboard test allowed at most 100 nonlinear updates and
stopped after five successive relative improvements below 0.02\% in the
unweighted traveltime RMS.  The maximum absolute $\log V_S$ increment was
limited to $\log(1.01)$, corresponding to a 1\% velocity-change cap.
After each regularizer was normalized on its control grid using the maximum
ratio of the diagonal of $\mathbf J_d^T\mathbf J_d$ to the corresponding
unit-regularization diagonal, the
relative levels were 0.1\% for diagonal damping and 5\% for both horizontal
and vertical second-order roughness, with uniform depth multipliers.

\begin{table}[t]
\centering
\caption{Reference values for the mantle branch of the USA elastic relations.
Values are linearly interpolated in depth; depths shallower than 33~km use
the first row.}
\label{tab:usa_mantle_reference}
\begin{tabular}{rccc}
\hline
Depth & $V_{S,\mathrm M}$ & $V_{P,\mathrm M}$ & $\rho_{\mathrm M}$ \\
(km) & (km~s$^{-1}$) & (km~s$^{-1}$) & (g~cm$^{-3}$) \\
\hline
33  & 4.360 & 7.760 & 3.320 \\
50  & 4.380 & 7.808 & 3.335 \\
100 & 4.450 & 7.950 & 3.380 \\
150 & 4.525 & 8.105 & 3.425 \\
\hline
\end{tabular}
\end{table}

We first conducted a synthetic recovery test using the same station--period
geometry, initial background, grids, and five control grids.  An initial
checkerboard with $\pm5\%$ $V_S$ anomalies, approximately
$3^{\circ}\times3^{\circ}$ horizontal lobes, and depth intervals of
0--20, 20--40, and 40--100~km was expressed as log-velocity anomalies relative
to the background.  These anomalies were fitted independently on each
control grid, and the five interpolated fields were averaged to define the
target.  Its velocity anomalies range from $-4.89\%$ to $+4.92\%$.
Synthetic traveltimes were generated from this target with the same $r=2$
forward calculation and perturbed by independent 1\% relative Gaussian
noise.  The test used unit data weights and the same regularization as the
real-data inversion.  It stopped after 11 updates, with RMS decreasing from
2.700 to 2.645~s.  The 99.5\% criterion selects update 9, whose RMS is
2.6453~s.

\begin{figure}[t]
  \centering
  \includegraphics[width=\linewidth]{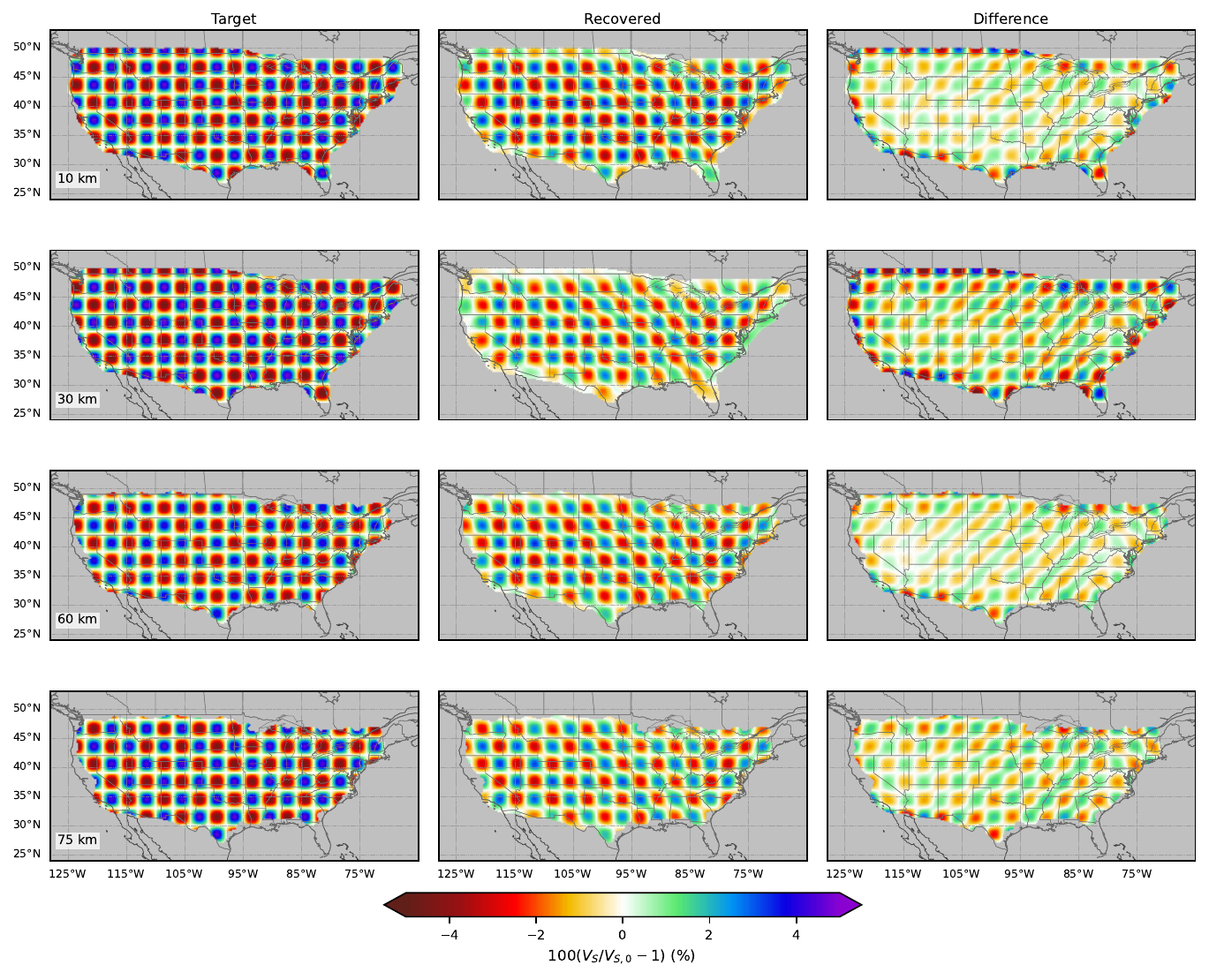}
  \caption{Synthetic checkerboard recovery using the contiguous-US station-period
  geometry at 10, 30, 60, and 75~km depth.  Columns show the target $V_S$
  perturbation, the model
  selected by the 99.5\% criterion (update 9), and recovered
  minus target.  The target uses the mean of five fitted control-grid
  log-velocity anomaly fields, with peak velocity perturbations of approximately
  $\pm4.9\%$.  Synthetic traveltimes contain 1\% relative Gaussian noise.
  Gray regions have values below 0.005 in the coverage proxy at the
  corresponding depth.  The proxy is obtained by interpolating, averaging, and normalizing
  the diagonals of $\mathbf J_d^T\mathbf J_d$ from the five control grids.
  Target and recovery show $100(V_S/V_{S,0}-1)$; differences are in
  percentage points.  All panels share the same color scale.}
  \label{fig:usa_checkerboard}
\end{figure}

The checkerboard geometry is recovered across much of the well-sampled region
at the displayed depths, with spatially variable amplitude and reduced
recovery toward the margins (Figure~\ref{fig:usa_checkerboard}).  Within the
displayed masks, correlations between the target and recovered percentage
anomalies range from 0.90 to 0.98 across the four slices.  A regression of
recovered against target anomalies through the origin gives amplitude
recovery of 49--70\%, with the weakest recovery at 30~km depth.

\begin{figure}[t]
  \centering
  \includegraphics[width=\linewidth]{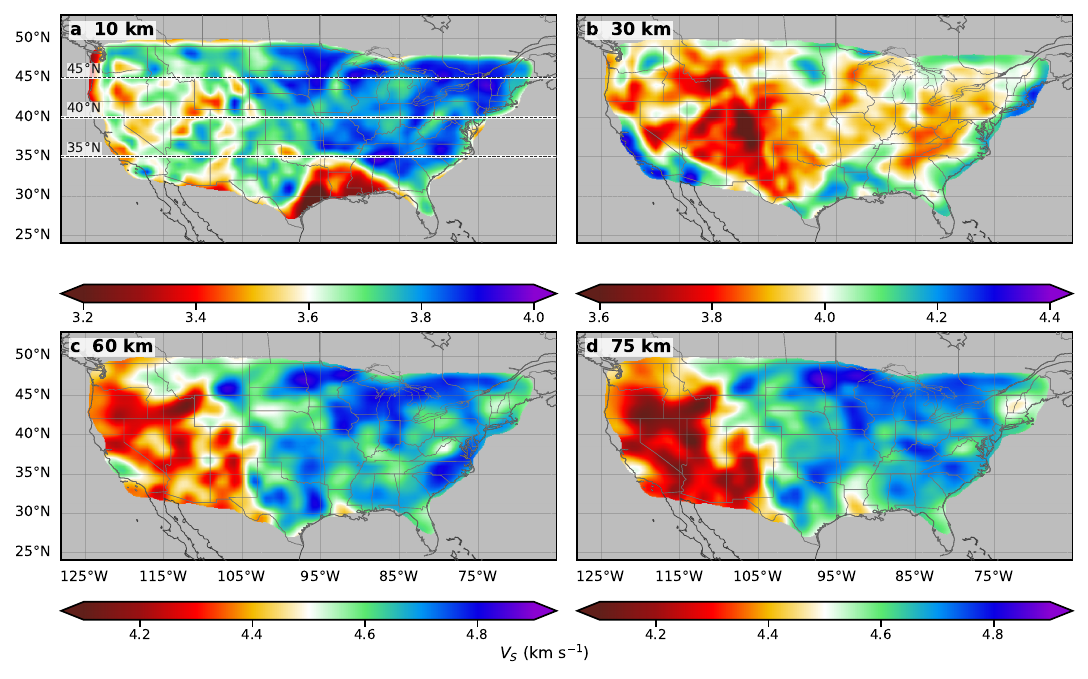}
  \caption{Absolute $V_S$ at 10, 30, 60, and 75~km depth in the contiguous-US
  model at update 20, the first satisfying the 99.5\% criterion for weighted RMS.
  Each depth uses an individual color
  range.  Gray regions have values below 0.005 in the coverage proxy averaged
  over depth.  Dashed lines in (a) locate the sections in
  Figure~\ref{fig:usa_profiles}.}
  \label{fig:usa_slices}
\end{figure}

\begin{figure}[t]
  \centering
  \includegraphics[width=0.90\linewidth]{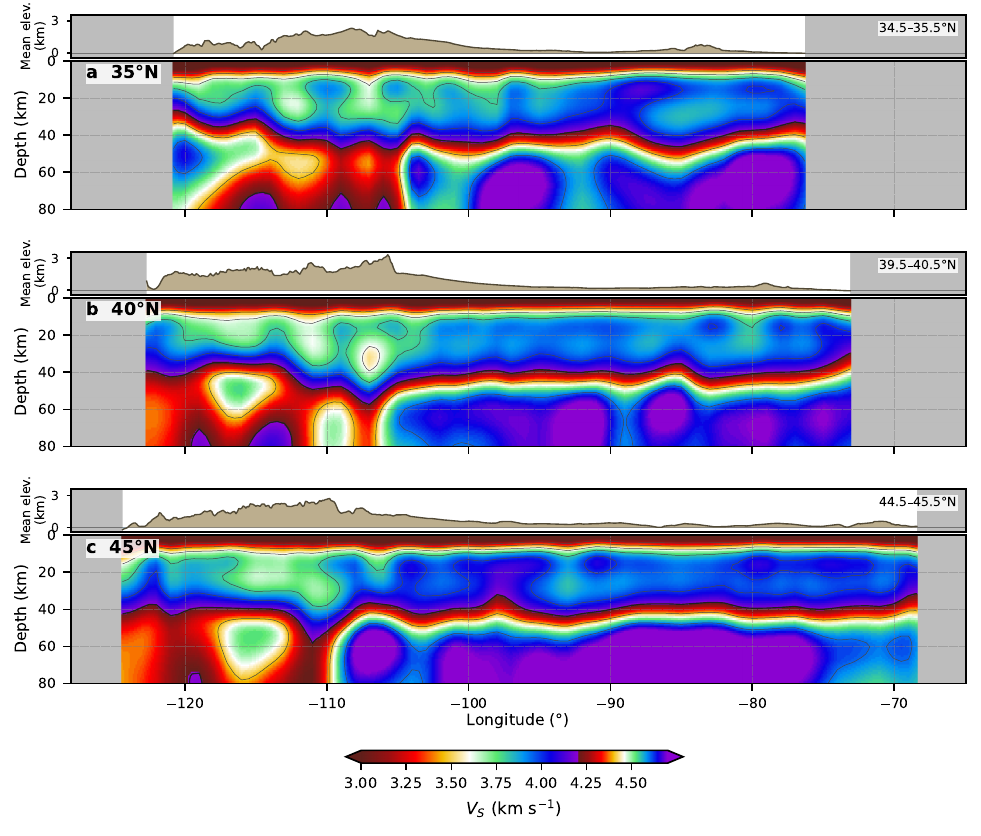}
  \caption{West--east sections of absolute $V_S$ along
  $35^{\circ}$N, $40^{\circ}$N, and $45^{\circ}$N through the model in
  Figure~\ref{fig:usa_slices}, to 80~km depth.  The same horizontal mask,
  based on the coverage proxy averaged over depth, is applied at all
  depths.  Contours run from 3.2 to 4.6~km~s$^{-1}$ at
  0.2~km~s$^{-1}$ intervals, with the 4.2~km~s$^{-1}$ contour emphasized.
  The color scale has separate segments below and above 4.2~km~s$^{-1}$.
  Above each section, mean elevation is averaged over a centered
  $1^{\circ}$ latitude band with area weighting.}
  \label{fig:usa_profiles}
\end{figure}

The real-data inversion stopped after 49 updates (4.6~h), reducing weighted
RMS from 8.411 to 1.519~s.  The 99.5\% criterion for weighted RMS
(equation~\eqref{eq:rms_decrease_attained}) selected update 20,
which was reached after 1.9~h of computation with a weighted RMS of 1.548~s.
The unweighted RMS decreased from 12.373 to 2.368~s over the complete run.
Figures~\ref{fig:usa_slices} and \ref{fig:usa_profiles} display the lateral
and depth-dependent variations in the selected real-data model.

The inverted model contains several first-order patterns recognized in previous
tomographic studies of the United States.  At 10~km depth, the lateral
variations broadly follow contrasts between sedimentary or tectonically
extended regions and higher-velocity regions in the continental interior.
At 60 and 75~km depth, a pronounced transition separates relatively low velocities
beneath the tectonically active western United States from higher velocities
beneath the central and eastern United States.  This transition persists in
the west--east sections and is consistent with earlier surface-wave models
\citep{BensenEtAl2009US,ShenRitzwoller2016}.

We also benchmarked one update using factorized and explicit Jacobians
at the model after 20 updates, with other settings fixed.  Cumulative operator storage
for the five augmented systems, counting shared factors once, was 21.75~GiB
with factorization and 528.24~GiB with explicit matrices, a factor of 24.29.
The explicit implementation also reused the shared factors to construct,
solve, and release one system at a time.  Its maximum simultaneously retained
operator storage was therefore 133.29~GiB, compared with 21.75~GiB for
factorization; these figures exclude temporary construction buffers.
Solving the five 20-step LSMR systems and combining their updates took
44.28 and 326.02~s, respectively, a factor of 7.36.  The resulting
$\log V_S$ updates differed by only $3.16\times10^{-8}$ in relative $L_2$ norm.

\clearpage


\section{Discussion and Conclusions}
\label{sec:discussion}

An aggregate adjoint efficiently gives one data-misfit gradient, whereas
obtaining individual traveltime kernels by repeating full-field adjoint
solves is expensive.  We derive the individual kernels by implicit
differentiation of the converged discrete Eikonal equations and restrict
each adjoint solve to the nodes contributing to that receiver, reusing the
gather's dependency graph, SCC decomposition, and cached block
factorizations.  Even with 1024 receivers, the forward solve and
construction of all individual kernels together took less than twice the
time for the forward solve and one aggregate discrete adjoint
(Figure~\ref{fig:receiver_scaling}).

Retaining the model and dispersion mappings as shared
factors reduces storage and the cost of repeated Jacobian and
transpose-Jacobian products.  In the U.S. single-update benchmark,
factorization reduced the maximum simultaneously retained operator storage
from 133.29 to 21.75~GiB and the time for the five LSMR solves and update
combination from 326.02 to 44.28~s.

The factorized representation also clarifies the relationship between the
present one-step formulation and conventional two-step surface-wave
tomography.  At the linearized-operator level, $\mathbf A$ represents
observation-specific lateral-propagation sensitivities, whereas
$\mathbf C\mathbf Q\mathbf K\mathbf D\mathbf P$ represents the shared
mapping from three-dimensional control-grid log-$V_S$ perturbations to
gather-layout, period-dependent phase slowness.  This mirrors the physical
decomposition exploited by two-step approaches, but without introducing an
intermediate inversion for period-specific velocity maps.  Instead, the two
components remain coupled within each Gauss--Newton update, allowing all
paths and periods to constrain the common three-dimensional model
simultaneously.  The separation therefore concerns reusable computation
and storage within a single coupled inverse problem.

The retained individual kernels provide the Jacobian products needed for
the regularized GN least-squares update.  In the synthetic comparison, the
reference GN workflow recovered
the target with fewer nonlinear updates and lower model error than
the L-BFGS and GD workflows at similar traveltime RMS values.

The U.S. inversion starts from a smoothed US.2016-derived model
(\ref{app:usa_initial_model}).  The projected checkerboard tests recovery
of approximately $3^{\circ}$ anomaly patterns
against the same background.  Their recovered geometry supports
interpretation of broad contrasts in well-sampled regions.
The inverted model shows lower velocities beneath the
tectonically active western United States and higher velocities beneath
the continental interior, consistent with earlier surface-wave models
\citep{BensenEtAl2009US,ShenRitzwoller2016}.

\newcommand{\manuscriptendstatements}{%
\acknowledgments
This work was supported by the Deep Earth Probe and Mineral Resources
Exploration National Science and Technology Major Project (Grant
2025ZD1009505).  Additional support was provided by the National Natural
Science Foundation of China (Grants 42504086, 42230806, and U23B6010) and the
China National Petroleum Corporation--Peking University Strategic Cooperation
Project of Fundamental Research.  We thank EarthScope Data Services, the
operators of the contributing seismic networks, and the investigators who
made the waveform data available.  We also thank Jing Chen (Nanyang
Technological University, Singapore) for helpful discussions.

\section*{Open Research Section}

The broadband waveform data and station metadata used in this study are
available from EarthScope Data Services (formerly the IRIS Data Management
Center) through the FDSN data-select and station services
\citep{TrabantEtAl2012}.  The service endpoint is
\url{https://service.earthscope.org}.  The tomography
implementation developed here extends the open-source SurfATT package
\citep{HaoEtAl2024,XuEtAl2025SurfATT}.  The interstation phase-velocity
dispersion curves derived in this study, the supporting synthetic and figure
data, and the source code implementing the factorized Gauss--Newton tomography
method developed here will be archived openly in Zenodo.  The associated data and software
DOIs and formal citations will be added before publication.


\section*{Inclusion in Global Research Statement}

This study used openly accessible seismic data and did not involve fieldwork,
human participants, Indigenous knowledge, biological samples, or direct
collaboration with local communities.  No permits or local authorizations
were required.  Authorship and acknowledgments recognize contributions in
accordance with AGU Publications criteria.

\section*{Conflict of Interest disclosure}

The authors declare there are no conflicts of interest for this manuscript.

}

\manuscriptendstatements

\bibliography{references}

\clearpage

\appendix

\section{Discrete Regularization Operators}
\label{app:regularization_discretization}

The regularization operators act on the $N_x\times N_y\times N_z$
physical grid.  The velocity field $\mathbf v$ is the physical $V_S$
increment for GN roughness and the current $V_S$ model for L-BFGS
roughness.  Its values $v_{ijk}$ are in km~s$^{-1}$, while $q_{ijk}$
denotes a dimensionless log-velocity increment for damping.
The fixed velocity scale is $V_*=4$~km~s$^{-1}$.
Longitude and latitude intervals are $\Delta\lambda$ and $\Delta\varphi$.
We set $c_j=\max(\cos\varphi_j,10^{-6})$, with latitude measured in radians.
All depth-dependent multipliers are unity in the reported inversions.

The horizontal operator stacks separate longitude and latitude rows:
\begin{align}
 (\mathbf R_x\mathbf v)_{ijk}
 &=\frac{1}{V_*}\sqrt{\frac{\Delta\varphi}{\Delta\lambda\,c_j}}
    (v_{i-1,j,k}-2v_{ijk}+v_{i+1,j,k}),
    &&2\leq i\leq N_x-1,
 \label{eq:regularization_horizontal_x}\\
 (\mathbf R_y\mathbf v)_{ijk}
 &=\frac{1}{V_*}\sqrt{\frac{c_j\,\Delta\lambda}{\Delta\varphi}}
    (v_{i,j-1,k}-2v_{ijk}+v_{i,j+1,k}),
    &&2\leq j\leq N_y-1,
 \label{eq:regularization_horizontal_y}\\
 \mathbf R_h\mathbf v&=
    [ (\mathbf R_x\mathbf v)^T, (\mathbf R_y\mathbf v)^T]^T.
 \label{eq:regularization_horizontal_stack}
\end{align}
The other horizontal index and the depth index span their full ranges.
These are second differences with angular aspect-ratio and latitude
weights; the displayed coefficients define the complete row scaling.
The horizontal sum gives each depth level unit quadrature weight.

For vertical roughness, depth is expressed numerically in kilometres,
$\zeta_k=z_k/(1~\mathrm{km})$.  Define
$h_-=\zeta_k-\zeta_{k-1}$, $h_+=\zeta_{k+1}-\zeta_k$, and
$\bar h_k=(h_-+h_+)/2$.  The nonuniform three-point row is
\begin{equation}
 (\mathbf R_v\mathbf v)_{ijk}
 =\frac{\sqrt{\bar h_k}}{V_*}\frac{2}{h_-+h_+}
   \left(\frac{v_{i,j,k+1}-v_{ijk}}{h_+}
         -\frac{v_{ijk}-v_{i,j,k-1}}{h_-}\right),
 \qquad 2\leq k\leq N_z-1.
 \label{eq:regularization_vertical}
\end{equation}
For uniform depth spacing $h$, its coefficients are
$[1,-2,1]/(V_*h^{3/2})$.  The factor $\sqrt{\bar h_k}$ supplies the local
depth-quadrature weight, with unit weight for each horizontal column.

The unit log-step damping operator is diagonal on every physical node:
\begin{equation}
 (\mathbf R_0\mathbf q)_{ijk}
   =\sqrt{\frac{\Delta\lambda_{\deg}\Delta\varphi_{\deg}}{0.25}}
     \,q_{ijk},
 \label{eq:regularization_log_damping}
\end{equation}
where $\Delta\lambda_{\deg}$ and $\Delta\varphi_{\deg}$ are the numerical
grid intervals in degrees; 0.25 is the reference area in square degrees.
On these uniform horizontal grids, $\mathbf R_0=a\mathbf I$, and the
scalar $a$ cancels under equation~\eqref{eq:regularization_scaling}.
The reference area is therefore only a scale convention before normalization.

Second-difference rows are included only for complete interior stencils.
Boundary nodes remain free and enter neighboring interior rows; no
additional boundary rows are introduced.

These rows enter the GN subproblem in equation~\eqref{eq:linearized_inverse},
with scales given by equation~\eqref{eq:regularization_scaling} for each
control grid.  The stated degree, kilometre, and km~s$^{-1}$ conventions
are used before this calibration.  For the diagonal estimates in
equation~\eqref{eq:regularization_scaling}, we average the component-wise
squares of eight transpose products with deterministic Rademacher probes
in each operator's range space; the damping diagonal is computed directly.

\section{Local Factored-Eikonal Updates and Derivatives}
\label{app:stencil}

This appendix gives the local update formulas and derivatives used to
construct $\mathbf U_{\tau,g}$ and $\mathbf U_{s,g}$ in
equation~\eqref{eq:active_matrices}.
For a fixed gather $g$, write $T=T_g$ and $s_f=s_{p_g,f}$, suppressing the
fixed gather and period indices.  Let $j$ denote the forward-grid node being updated and
$h_x,h_y$ the forward-grid spacings in local surface coordinates $(x,y)$.
At node $j$, write the inverse metric as
$\mathbf{M}_j=\left[\begin{smallmatrix}a_j&-c_j\\-c_j&b_j\end{smallmatrix}\right]$,
which is positive definite.
The Eikonal equation used by the numerical candidate update is
\begin{equation}
  a_j T_x^2+b_j T_y^2-2c_j T_xT_y=s_{f,j}^2,
  \qquad a_j>0,\quad a_j b_j-c_j^2>0.
  \label{eq:generalized_eikonal_appendix}
\end{equation}
Here $a_j$, $b_j$, and $c_j$ are fixed nodal metric coefficients, and
$s_{f,j}$ is the forward-grid slowness; for a flat surface in Cartesian
coordinates, $a_j=b_j=1$ and $c_j=0$.  After substituting $T=T_0\tau$, define
$T_{0x,j}=(\partial T_0/\partial x)_j$ and
$T_{0y,j}=(\partial T_0/\partial y)_j$.  We let
$\sigma_x,\sigma_y\in\{-1,1\}$ identify the selected neighbor side in each
coordinate direction and define
$j_x=j+\sigma_x\mathbf{e}_x$ and $j_y=j+\sigma_y\mathbf{e}_y$, where
$\mathbf{e}_x$ and $\mathbf{e}_y$ are the unit index increments in the two
coordinate directions.  The active finite differences are
\begin{align}
  u_x &\equiv T_x
  \simeq T_{0x,j}\tau_j
  +T_{0,j}\frac{\sigma_x(\tau_{j_x}-\tau_j)}{h_x}
  =\eta_x\tau_j+\zeta_x, \\
  \eta_x &=T_{0x,j}-\frac{\sigma_x T_{0,j}}{h_x},
  \qquad
  \zeta_x=\frac{\sigma_x T_{0,j}}{h_x}\tau_{j_x},
  \label{eq:factored_x_stencil}\\
  u_y &\equiv T_y
  \simeq T_{0y,j}\tau_j
  +T_{0,j}\frac{\sigma_y(\tau_{j_y}-\tau_j)}{h_y}
  =\eta_y\tau_j+\zeta_y, \\
  \eta_y &=T_{0y,j}-\frac{\sigma_y T_{0,j}}{h_y},
  \qquad
  \zeta_y=\frac{\sigma_y T_{0,j}}{h_y}\tau_{j_y}.
  \label{eq:factored_y_stencil}
\end{align}
For an update that uses both coordinate directions, trial values of $\tau_j$
are obtained from the roots of
\begin{equation}
  G_j(\tau_j)
  =a_j u_x^2+b_j u_y^2-2c_j u_xu_y-s_{f,j}^2=0.
  \label{eq:active_quadratic}
\end{equation}
If only one coordinate is admissible, the reduced candidates satisfy
\begin{align}
  G_j^{(x)}&=u_x-\varepsilon s_{f,j}
  \sqrt{\frac{b_j}{a_j b_j-c_j^2}}=0,
  \label{eq:axis_x_candidate}\\
  G_j^{(y)}&=u_y-\varepsilon s_{f,j}
  \sqrt{\frac{a_j}{a_j b_j-c_j^2}}=0,
  \label{eq:axis_candidates}
\end{align}
where $\varepsilon\in\{-1,1\}$ denotes the tested root branch.  The forward
solver evaluates the real roots of these local equations, retains those that
satisfy the positivity and upwind-causality conditions, and selects the
smallest admissible trial value.  After convergence, the candidate nearest the
stored $\tau_j$ is reconstructed and its analytic derivatives are cached.
We denote this selected nodal update by $\phi_j$ ($\phi_{g,j}$ in the full
notation); it depends on the active neighboring values of $\tau$ and on
the forward-grid slowness.  Collecting these updates over free nodes gives
$\boldsymbol{\phi}_g$ in equation~\eqref{eq:discrete_fixed_point}.
The source-neighborhood values are prescribed as $\tau_j=1$ and have zero
variation.  All remaining nodes are free nodes; their values form
$\boldsymbol{\tau}_g$, and only their local equations contribute rows to
$\mathbf{B}_g$.

For the branch that uses both coordinate directions, define
\begin{equation}
  \gamma_j=\frac{\partial G_j}{\partial\tau_j}
  =2\left[
    (a_j u_x-c_j u_y)\eta_x+(b_j u_y-c_j u_x)\eta_y
  \right].
  \label{eq:active_denominator}
\end{equation}
Holding the source-factor quantities $T_0$, $T_{0x}$, and $T_{0y}$ fixed,
the nonzero neighbor and local-slowness derivatives of the selected update
$\tau_j=\phi_j$ are
\begin{align}
  \frac{\partial\phi_j}{\partial\tau_{j_x}}
  &=-\frac{2(a_j u_x-c_j u_y)}{\gamma_j}
  \frac{\sigma_x T_{0,j}}{h_x},
  \label{eq:active_neighbor_x_derivative}\\
  \frac{\partial\phi_j}{\partial\tau_{j_y}}
  &=-\frac{2(b_j u_y-c_j u_x)}{\gamma_j}
  \frac{\sigma_y T_{0,j}}{h_y},\\
  \frac{\partial\phi_j}{\partial s_{f,j}}
  &=\frac{2s_{f,j}}{\gamma_j}.
  \label{eq:active_neighbor_derivatives}
\end{align}
The coefficients for free neighbors populate $\mathbf U_{\tau,g}$, while
the local-slowness coefficient contributes to $\mathbf U_{s,g}$.
Source slowness is bilinearly interpolated from the four surrounding
forward-grid nodes and enters $T_0$ and its coordinate derivatives.
For an update using both coordinate directions, the source-factor
derivatives are
\begin{align}
  \frac{\partial\phi_j}{\partial T_{0x,j}}
  &=-\frac{2(a_j u_x-c_j u_y)\tau_j}{\gamma_j},
  \label{eq:T0x_derivative}\\
  \frac{\partial\phi_j}{\partial T_{0y,j}}
  &=-\frac{2(b_j u_y-c_j u_x)\tau_j}{\gamma_j},\\
  \frac{\partial\phi_j}{\partial T_{0,j}}
  &=-\frac{2}{\gamma_j}\left[
    (a_j u_x-c_j u_y)
      \frac{\sigma_x(\tau_{j_x}-\tau_j)}{h_x}
    +(b_j u_y-c_j u_x)
      \frac{\sigma_y(\tau_{j_y}-\tau_j)}{h_y}
  \right].
  \label{eq:T0_derivatives}
\end{align}
The corresponding single-coordinate derivatives follow from
equations~\eqref{eq:axis_x_candidate}--\eqref{eq:axis_candidates}.

The chain rule through $T_0$, $T_{0x}$, $T_{0y}$, and the source
interpolation weights $\boldsymbol\chi_{s,g}$ supplies the source-node
contributions to $\mathbf U_{s,g}$.  The separate direct receiver term is
given in equation~\eqref{eq:receiver_derivatives}.
For a windowed gather, the locally computed sensitivity is returned by
transpose interpolation and assembled in the global phase-velocity layout
with zero contributions outside the window, as in
equation~\eqref{eq:individual_row}.

Equations~\eqref{eq:active_neighbor_x_derivative}--\eqref{eq:T0_derivatives}
apply with the selected two-coordinate branch held fixed and
$|\gamma_j|$ bounded away from zero.  The single-coordinate formulas
have the same requirement on their local denominators.  At an exact tie,
a unique classical derivative need not exist; the fixed candidate order
determines which branch is differentiated.

\section{Cyclic Active-Stencil Graphs}
\label{app:cycles}

The SCC decomposition in Section~\ref{sec:graph_kernel_evaluation} is
obtained with the Kosaraju--Sharir algorithm \citep{SedgewickWayne2011}.
Within each receiver's ancestor closure, the blocks are processed in
adjoint order, from the receiver toward the source.  A cyclic component
$\mathcal C$ is solved as
\begin{equation}
  \mathbf{B}_{g,\mathcal{C}\mathcal{C}}^T
  \boldsymbol{\lambda}_{i,\mathcal{C}}
  =\mathbf{h}_{i,\mathcal{C}}
   -\sum_{\mathcal{D}\ne\mathcal{C}}
    \mathbf{B}_{g,\mathcal{D}\mathcal{C}}^T
    \boldsymbol{\lambda}_{i,\mathcal{D}}.
  \label{eq:scc_block_solve}
\end{equation}
Nonzero terms in the sum come from components $\mathcal D$ that have
already been processed.

Cyclic SCCs containing at most 64 nodes are solved with cached pivoted LU
factors.  A larger or numerically unsupported block is solved on the full
ancestor subgraph by BiCGSTAB \citep{VanDerVorst1992}; restarted GMRES with a
restart length of 40 is used if BiCGSTAB does not converge or encounters a
breakdown \citep{SaadSchultz1986}.  In the synthetic inversions, these
iterative solves require the linear residual $\mathbf{r}_{\mathrm{lin}}$ to satisfy
$\|\mathbf{r}_{\mathrm{lin}}\|_2
\leq\max(10^{-12},10^{-10}\|\mathbf{b}\|_2)$, where $\mathbf{b}$ is the
restricted right-hand side.

\clearpage
\section{Propagation Timing Components}
\label{app:propagation_timing}

Table~\ref{tab:propagation_timing} lists the timing components for 1024
receivers, with measurement boundaries given in the caption of
Figure~\ref{fig:receiver_scaling}.

\begin{table}[!hbp]
\centering
\caption{Propagation-stage times for 1024 receivers in Figure~\ref{fig:receiver_scaling},
in milliseconds.  Entries are medians of 15 repetitions; totals use
per-repetition stage sums.  The discrete aggregate routine includes SCC
preparation.  Packing retained rows into CSR storage is excluded.}
\label{tab:propagation_timing}
\small
\begin{tabular}{p{0.57\linewidth}rr}
\hline
Stage or workload & $100\times100$ & $200\times200$ \\
\hline
Common forward solve & 32.35 & 129.96 \\
Active-stencil reconstruction & 2.113 & 8.516 \\
Individual workspace/SCC preparation & 0.271 & 1.155 \\
All individual discrete kernels & 29.07 & 105.38 \\
Discrete aggregate routine & 0.847 & 3.696 \\
\hline
Discrete total (plotted) & 2.964 & 12.205 \\
Discrete individual (plotted) & 31.46 & 115.04 \\
Continuous total (plotted) & 1.487 & 6.232 \\
Continuous individual (plotted) & 1358.68 & 5652.65 \\
\hline
\end{tabular}
\end{table}

\section{Synthetic Experiment Configuration and Additional Regularization Sweeps}
\label{app:optimizer_parameters}

The synthetic background $\beta_0(z)$ in
Section~\ref{sec:small_anomaly_results} is horizontally uniform and piecewise
linear in $V_S$ through the depth--velocity pairs
$(0,3.0)$, $(10,3.2)$, $(20,3.5)$, $(35,3.8)$, $(60,4.1)$, and
$(100,4.35)$, with depth in km and velocity in km~s$^{-1}$.
Its physical-grid samples define the common initial model
$\boldsymbol{\beta}_0$.
The physical domain spans $100$--$110^{\circ}$E, $20$--$30^{\circ}$N, and
0--100~km.  The $18\times18\times22$ control grid spans
$100$--$110.2^{\circ}$E and $20$--$30.2^{\circ}$N at $0.6^{\circ}$ intervals,
and $-2.5$--$102.5$~km at 5~km intervals.  The target log-velocity
perturbations are prescribed on this grid as
\begin{equation}
  \begin{aligned}
  u^*_{abc}&=a_*\,
    \sin\!\left[\frac{\pi(x_a-100^{\circ})}{1.8^{\circ}}\right]
    \sin\!\left[\frac{\pi(y_b-20^{\circ})}{1.8^{\circ}}\right]
    \sin(2\pi q_c),\\
  q_c&=\frac{\log\beta_0(\widehat z_c)-\log\beta_0(0)}
            {\log\beta_0(100)-\log\beta_0(0)},
  \end{aligned}
  \label{eq:synthetic_target_controls}
\end{equation}
where $\widehat z_c$ clips each control depth to 0--100~km.
Trilinear interpolation gives the target
$\beta_*(\mathbf{x},z)=\beta_0(z)\exp[(\mathbf{P}\mathbf{u}^*)(\mathbf{x},z)]$.
The amplitude $a_*$ is normalized so that
$\max|\mathbf{P}\mathbf{u}^*|=0.05$ on the physical grid.
Independent Gaussian traveltime perturbations have standard deviation
$0.005\,d_i^{\mathrm{true}}$.

The Eikonal convergence tolerance was $10^{-4}$~s for both the mean and
maximum nodal traveltime changes over a sweeping cycle.
The fixed source neighborhood had
$|x-x_s|\leq h_x$ and $|y-y_s|\leq h_y$, where $h_x,h_y$ are the horizontal
grid intervals on the forward grid.

Table~\ref{tab:optimizer_parameters} lists all configurations in the formal
comparison.

\begin{table}[!hbp]
\centering
\caption{Regularization parameters for the nine formal synthetic inversions.
For GN, the Horizontal, Vertical, and Damping entries are the dimensionless
$\theta_\ell$ ratios in equation~\eqref{eq:regularization_scaling}.
For L-BFGS, the Horizontal and Vertical entries are the cumulative roughness
weights $w_\ell$; the two sets of values are therefore not directly comparable.
GN roughness acts on the physical $V_S$ increment, while its damping acts on
the physical $\log V_S$ increment.  The $r=2$ GN case uses observations
synthesized on its refined forward grid.}
\label{tab:optimizer_parameters}
\small
\begin{tabular}{lccccc}
\hline
Method & Label & Target & Horizontal & Vertical & Damping \\
\hline
GN ($r=1$) & $0.25\times$ & update & 0.0125 & 0.050 & 0.005 \\
GN ($r=1$) & $1\times$ & update & 0.0500 & 0.200 & 0.005 \\
GN ($r=2$) & $1\times$ & update & 0.0500 & 0.200 & 0.005 \\
GN ($r=1$) & $4\times$ & update & 0.2000 & 0.800 & 0.005 \\
L-BFGS & $0\times$ & absolute $V_S$ & 0 & 0 & 0 \\
L-BFGS & $0.25\times$ & absolute $V_S$ & $5.0\times10^{-7}$ & $2.5\times10^{-7}$ & 0 \\
L-BFGS & $1\times$ & absolute $V_S$ & $2.0\times10^{-6}$ & $1.0\times10^{-6}$ & 0 \\
L-BFGS & $4\times$ & absolute $V_S$ & $8.0\times10^{-6}$ & $4.0\times10^{-6}$ & 0 \\
GD & $0\times$ & none & 0 & 0 & 0 \\
\hline
\end{tabular}
\end{table}

L-BFGS and GD retain cumulative control coefficients
$\mathbf m^{\mathrm{cum}}$ on a fixed control grid, representing the current
model as
$\boldsymbol{\beta}=\boldsymbol{\beta}_0\odot
\exp(\mathbf P\mathbf m^{\mathrm{cum}})$.
Their local increments follow equation~\eqref{eq:control_increment_mapping}.

For L-BFGS, the data objective is normalized by its fixed initial value and
supplemented by the cumulative roughness penalty
$\tfrac12\sum_{\ell\in\{h,v\}}w_\ell
\|\mathbf R_\ell\boldsymbol\beta\|_2^2$,
using the weights in Table~\ref{tab:optimizer_parameters}.
The discrete forms and units of $\mathbf R_h$ and $\mathbf R_v$ are given in
\ref{app:regularization_discretization}.  The physical-grid velocity gradient
of these penalties is mapped to control space by $\mathbf P^T\mathbf D$,
with $\mathbf D=\operatorname{diag}(\boldsymbol\beta)$ evaluated at the
current model.

LSMR used at most 20 iterations per nonlinear update, with relative stopping
parameters $\mathrm{atol}=\mathrm{btol}=10^{-6}$ and a condition limit of $10^{10}$.
The bound in equation~\eqref{eq:bounded_trial_update} was $b=0.02$.
Backtracking reduced $\alpha$ by a factor of 0.6 and allowed at most ten trials.
GN retained the trial with the lowest data
misfit, stopping the search when a smaller step no longer improved the best
decreasing trial.

L-BFGS retained up to five valid curvature pairs in control space; GD used
the negative control gradient.  Both accepted the first trial that did not
increase their data-plus-roughness objective.  Only accepted trials updated
the model and the L-BFGS history.  A zero gradient or failure to
accept any trial terminated the run.  Cumulative inversion times include
rejected trials; the independent terminal forward check and plotting were
timed separately.

In the three-method validation, the aggregate data gradient through the
complete model-to-data chain and
$\mathbf{J}^T\mathbf{W}\mathbf{e}$ agreed to a relative error of
$1.1\times10^{-15}$ under a common normalization.  Transpose and
directional-derivative tests verified the control mapping and regularized
objective.

The regularization search comprised 18 runs at $r=1$.
Three GN runs tested $(\theta_h,\theta_v)$ pairs of $(0.025,0.10)$,
$(0.05,0.20)$, and $(0.10,0.40)$, with at most 40 updates per run and
damping fixed at 0.005.  The six-case L-BFGS and GD sweeps used multipliers
$0$, $10^{-3}$, $10^{-2}$, $10^{-1}$, $1$, and $10$ relative to cumulative
horizontal and vertical base weights of $2\times10^{-4}$ and
$1\times10^{-4}$, respectively, with limits of 80 L-BFGS or 120 GD updates.
A finer L-BFGS sweep used 0.005, 0.01, and 0.02 times the base
weights with a limit of 120 updates each.

\section{Initial Model for the Contiguous-US Application}
\label{app:usa_initial_model}

\begin{figure}[!hbp]
 \centering
 \includegraphics[width=\linewidth]{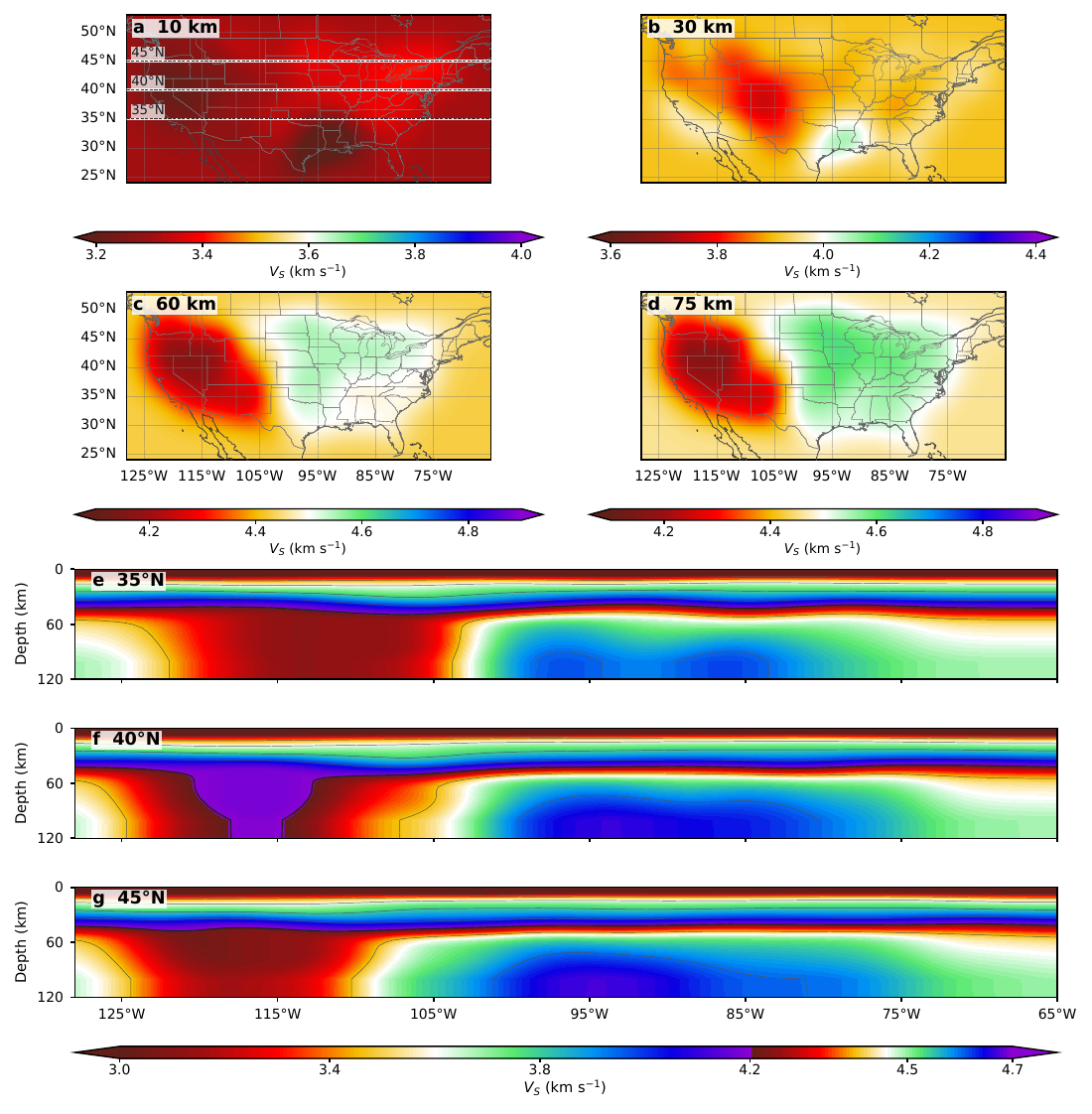}
 \caption{Initial $V_S$ model for the contiguous-US inversion.
 (a--d) Horizontal slices at 10, 30, 60, and 75~km, with the same
 depth-specific color limits as Figure~\ref{fig:usa_slices}.
 (e--g) Sections at 35, 40, and 45$^{\circ}$N, using the color scale of
 Figure~\ref{fig:usa_profiles} and extending to the 120~km model bottom.
 The complete initial volume is shown without a data-coverage mask.
 Dashed lines in (a) locate the vertical sections.}
 \label{fig:usa_initial_model}
\end{figure}

Figure~\ref{fig:usa_initial_model} shows the initial $V_S$ model prepared
from US.2016 \citep{ShenRitzwoller2016}.  Missing cells and lateral padding
were filled with the valid mean at each depth before Gaussian smoothing
with standard deviations of $2.5^{\circ}$ horizontally and 7.5~km vertically.
The smoothed profiles were adjusted to increase monotonically with depth
and extended linearly to 120~km.

\end{document}